\documentclass[11pt,urlcolor=blue, linkcolor=blue]{article} 
\usepackage{cite}

\usepackage{amsmath, amsthm, amssymb,slashed}
\usepackage{ifpdf}
\ifpdf
  \usepackage[pdftex]{graphicx}
  \usepackage{epstopdf}
\else
  \usepackage[dvips]{graphicx}
\fi
\parskip=\baselineskip

\allowdisplaybreaks[1]
\numberwithin{equation}{section}

\newcommand{\cblue}[1]{\textcolor{black}{#1}}

\usepackage{upgreek}

\newcommand{\GSD}{\text{GSD}}
\usepackage{tabularx}

\DeclareMathAlphabet{\mathpzc}{OT1}{pzc}{m}{it}

\newcommand{\bea}{\begin{eqnarray}}
\newcommand{\eea}{\end{eqnarray}}
\def\be{\begin{equation}}
\def\ee{\end{equation}}

\usepackage{color}
\usepackage[colorlinks,citecolor=blue]{hyperref}
\definecolor{red}{rgb}{1,0,0}
\definecolor{blue}{rgb}{0,0,1}
\definecolor{dblue}{rgb}{0,0,0.4}
\definecolor{green}{rgb}{0,1,0}
\definecolor{black}{rgb}{0,0,0}
\definecolor{white}{rgb}{1,1,1}

\definecolor{brn}{rgb}{.8,.4,.0}
\definecolor{redo}{rgb}{1,.5,.0}
\definecolor{ddgrn}{rgb}{0,0.4,0}
\definecolor{dgrn}{rgb}{0,0.55,0}
\definecolor{dbl}{rgb}{0,0,0.5}

\usepackage[bbgreekl]{mathbbol}
\usepackage{amscd}

\newcommand{\Z}{\mathbb{Z}}

\newcommand{\R}{\mathbb{R}}

\newcommand{\ii}{\hspace{1pt}\mathrm{i}\hspace{1pt}}
\newcommand{\dd}{\hspace{1pt}\mathrm{d}}
\newcommand{\<}{\langle} 
\renewcommand{\>}{\rangle}

\newcommand{\eq}[1]{(\ref{#1})} 
\newcommand{\eqn}[1]{Eq.~(\ref{#1})}

\newcommand{\prt}{\partial}

\newcommand{\bpm}{\begin{pmatrix}}
\newcommand{\epm}{\end{pmatrix}}
\newcommand{\bmm}{\begin{matrix}}
\newcommand{\emm}{\end{matrix}}

\newcommand{\cB}{ {\cal B} }

\newcommand{\cH}{ {\cal H} }

\newcommand{\cS}{ {\cal S} } 
\newcommand{\cT}{ {\cal T} }

\newcommand{\al}{\alpha} 
 
\newcommand{\del}{\delta}

\renewcommand{\th}{\theta} 
 
\newcommand{\si}{\sigma}

\def\Z{{\mathbb{Z}}}

\def\R{{\mathbb{R}}}

\def\cM{{\cal M}}

\DeclareRobustCommand\clefG{\includegraphics[height=3.95ex]{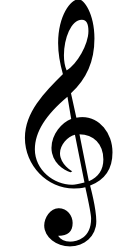}}
\DeclareRobustCommand\clefF{\includegraphics[height=3ex]{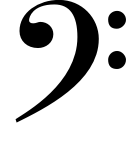}}
\DeclareRobustCommand\clefC{\includegraphics[height=2.6ex]{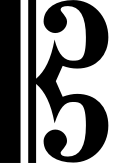}}

\newcommand{\Gfootnote}[1]{%
\let\oldthefootnote=\thefootnote%
\stepcounter{mpfootnote}%
\addtocounter{footnote}{-1}%
\renewcommand{\thefootnote}{\clefG}
\footnote{#1}%
\let\thefootnote=\oldthefootnote%
}

\newcommand{\Ffootnote}[1]{%
\let\oldthefootnote=\thefootnote%
\stepcounter{mpfootnote}%
\addtocounter{footnote}{-1}%
\renewcommand{\thefootnote}{\clefF}
\footnote{#1}%
\let\thefootnote=\oldthefootnote%
}

\newcommand{\Cfootnote}[1]{%
\let\oldthefootnote=\thefootnote%
\stepcounter{mpfootnote}%
\addtocounter{footnote}{-1}%
\renewcommand{\thefootnote}{\clefC}
\footnote{#1}%
\let\thefootnote=\oldthefootnote%
}

\usepackage{enumitem,xcolor,cleveref}
\newcommand{\nn}{\nonumber}

\newcommand{\MCG}{\mathrm{MCG}}
\newcommand{\cF}{\mathcal{F}}
\newcommand{\cN}{\mathcal{N}}
\newcommand{\SL}{\text{SL}}

\newcommand{\tL}{\mathrm{L}}

\usepackage[letterpaper]{geometry}
\usepackage{anysize}\marginsize{20mm}{20mm}{20mm}{20mm}

\usepackage{comment}

\def \- {\!\smallsetminus\!}

\begin{document}
\begin{titlepage}
\begin{flushright}
\end{flushright}
\vskip 1.0in
\begin{center}

{\bf\LARGE{Quantum Statistics and Spacetime Topology:\\[5.5mm] 
Quantum Surgery Formulas }}


\vskip0.5cm 
\Large{Juven Wang$^{1,2}$,{\Gfootnote{e-mail: {\tt juven@ias.edu}} }
Xiao-Gang Wen$^3$,{\Cfootnote{e-mail: {\tt xgwen@mit.edu}}} 
Shing-Tung Yau$^{4,2,5}${\Ffootnote{e-mail: {\tt {yau@math.harvard.edu}}}}
} 
\vskip.5cm
  {\small{\textit{$^1$School of Natural Sciences, Einstein Drive, Institute for Advanced Study, Princeton, NJ 08540, USA}\\}}
 \vskip.2cm
 {\small{\textit{$^2${Center of Mathematical Sciences and Applications, Harvard University,  Cambridge, MA 02138, USA} \\}}
}
 \vskip.2cm
 {\small{\textit{$^3$Department of Physics, Massachusetts Institute of Technology, Cambridge, MA 02139, USA \\}}
}
\vskip.2cm
{\small{\textit{$^4${Department of Mathematics, Harvard University,  Cambridge, MA 02138, USA} \\}}
}
\vskip.2cm
{\small{\textit{$^5${Department of Physics, Harvard University,  Cambridge, MA 02138, USA} \\}}
}


\end{center}
\vskip.5cm
\baselineskip 16pt
\begin{abstract}

To formulate the universal constraints of quantum statistics data of generic long-range entangled quantum systems, 
we introduce the geometric-topology surgery theory on spacetime manifolds where quantum systems reside, 
cutting and gluing the associated quantum amplitudes,
specifically in 2+1 and 3+1 spacetime dimensions.
First, we introduce the fusion data for worldline and worldsheet operators capable of creating  
anyonic excitations of  particles and strings, well-defined in gapped states of matter with intrinsic topological orders.
Second, we introduce the braiding statistics data of particles and strings, such as the geometric Berry matrices for
particle-string Aharonov-Bohm, 3-string, 4-string, or multi-string adiabatic loop braiding process, 
encoded by submanifold linkings, in the closed spacetime 3-manifolds and 4-manifolds.
Third, we derive  new ``quantum surgery''  formulas and constraints, analogous to Verlinde formula associating 
fusion and braiding statistics data via spacetime surgery, essential for defining
the theory of topological orders, 3d and 4d TQFTs and potentially correlated to bootstrap boundary physics such as gapless modes, extended defects,
2d and 3d conformal field theories or quantum anomalies.\\

This article is meant to be an extended and further detailed elaboration of 
 our previous work \cite{Wang2016qhf1602.05951}
 and
 Chapter 6 of Ref.~\cite{JWangthesis}. 
Our theory applies to general quantum theories and quantum mechanical systems, also applicable to, but not necessarily requiring the quantum field theory description.


\end{abstract}
\end{titlepage}

\tableofcontents   


\section{Introduction}

Geometry and topology have long perceived to be intricately related to our understanding of the physics world. Our physical Universe is known to be quantum in nature. For a system with many-body quantum degrees of freedom, we have a well-motivated purpose of determining the governing laws of quantum statistics. Quantum statistics, to some extent,  behave as hidden long-range ``forces'' or ``interactions'' between the quantum quasi-excitations. One of the goals of our present work is formulating the 
\emph{quantum version} of
constraints from geometric topology and surgery properties of spacetimes (as manifolds, either smooth differentiable, or triangulable on a lattice) in order 
to develop theoretical equations governing the quantum statistics of general quantum theories.

\begin{figure}[!h]
\centering
{
\includegraphics[scale=0.45]{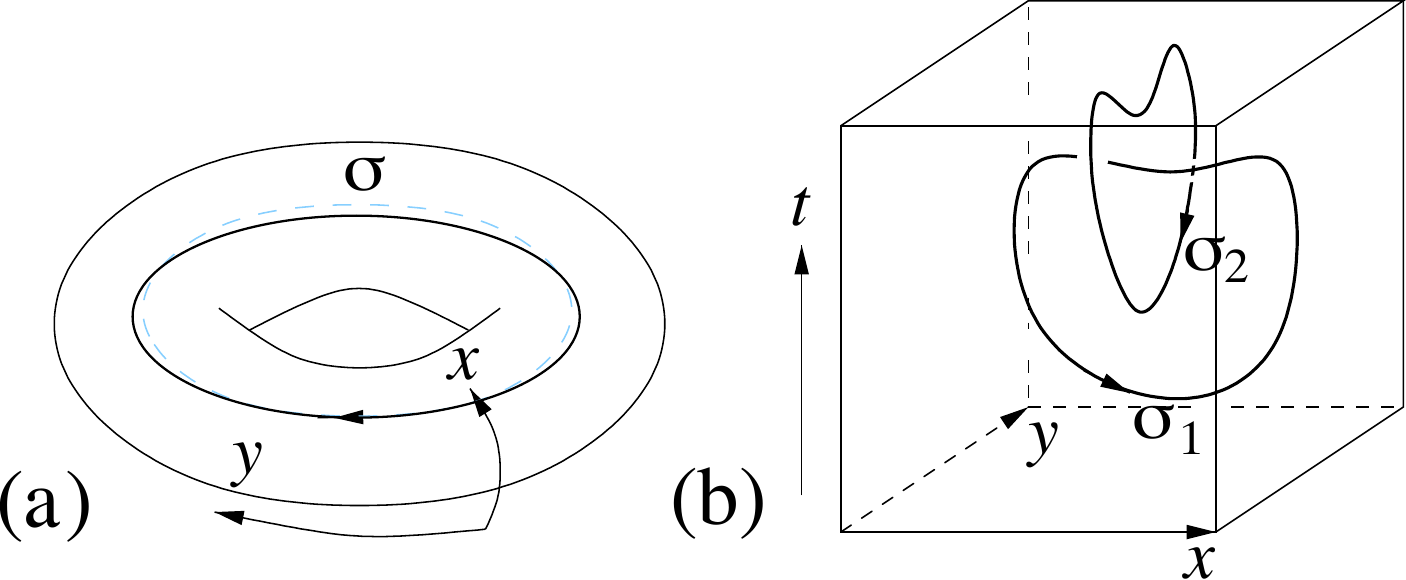}
}
\caption{
(a) A topologically-ordered ground state on a spatial 2-torus $T^2_{xy}$ is labeled by a quasiparticle $\si$. 
(b) The quantum amplitude of two linked 
\cblue{spacetime trajectories of anyons $\si_1$ and $\si_2$} in 2+1D
is proportional to a complex number $\cS_{\si_1 \si_2}$, which is related to the modular SL(2,$\Z$) data to be introduced later. 
}
\label{lnklp}
\end{figure}

The 
fractional quantum Hall effect was discovered decades ago \cite{TSG8259}.
A quantum theory of the wavefunction of fractional quantum Hall effect is formulated subsequently \cite{L8395}.
The intrinsic relation between the topological quantum field theories (TQFT) and the topology of manifolds
was found 
years after \cite{Schwarz:1978cn, Witten:1988hf}. 
These breakthroughs  
partially 
motivated the study of topological order \cite{Wenrig} as a new state of matter in quantum many-body systems
and in condensed matter systems \cite{Wen:2012hm}. 
Topological orders are defined as the gapped states of matter with physical properties depending on 
global 
topology (such as the ground state degeneracy (GSD)), 
robust against any local perturbation and any symmetry-breaking perturbation.
Accordingly, topological orders cannot be characterized by 
the old paradigm of symmetry-breaking 
phases of matter via the Ginzburg-Landau theory \cite{GL5064,LanL58}.
The systematic studies 
of 2+1 dimensional spacetime\footnote{For abbreviation, we write $n+1$D for an $n$+1-dimensional spacetime. 
We write $m$d for an $m$-dimensional spacetime.
We write $n$D simply for the $n$-dimensional manifold or $n$D space.
} (2+1D) 
topological orders enhance our understanding of the real-world plethora phases including
quantum Hall states and spin liquids \cite{BalentsSL}.
In this work, we explore the 
constraints between the 2+1D and 3+1D 
topological orders
and the geometric-topology properties of 3- 
and 4-manifolds. 
\cblue{We only focus on 2+1D / 3+1D topological orders 
with 
GSD insensitive to the system size
and
with a finite number of types of topological excitations 
creatable from 1D line and 2D surface operators.
}
Specifically, the open ends of 1D line operators give rise to quasi-excitations of anyons.
The open ends of 2D surface operators give rise to quasi-excitations of anyonic strings.
Conversely, the worldlines of anyons become 1D line operators (see Fig.\ref{lnklp}), while
the worldsheets of anyonic strings become 2D surface operators.

\cblue{In this work, we mainly apply the tools of
quantum mechanics in physics and 
surgery theory in mathematics \cite{thurston1997three, gompf19994}.}
Our main results are:
(1) We provide the \emph{fusion data} for worldline and worldsheet operators creating excitations of particles (i.e. anyons \cite{Wilczek:1990ik}) and strings (i.e. anyonic strings) in topological orders.
(2) We provide the \emph{braiding statistics data} of particles and strings encoded by submanifold linking, in the 3- and 4-dimensional closed spacetime manifolds.
(3) By ``cutting and gluing'' (or ``cutting and sewing'' in synonym) quantum amplitudes,
we derive constraints between the fusion and braiding statistics data analogous to Verlinde formula 
\cite{Verlinde:1988sn, Moore:1988qv, Witten:1988hf}  
for 
2+1 and 3+1D topological orders.


\section{Quantum Statistics: Fusion and Braiding Statistics Data}
Imagine a renormalization-group-fixed-point 
topologically ordered quantum system 
on a spacetime manifold $\cM$. The 
manifold can be viewed as a long-wavelength continuous limit of certain lattice regularization of the system.\footnote{Mathematically, it is proven that
all smooth thus differentiable manifolds are always triangulable. This can be proven via Morse theory, which implies that we 
only need to show the triangulation of piecewise-linear (PL) handle-attachments.
However, the converse statement may not be true in general. 
Thus here
we focus on going from 
the smooth thus differentiable manifolds (the continuum) to the triangulable manifolds (the discrete).
} 
We aim to compute the quantum amplitude from ``gluing'' one ket-state $| R\rangle$ with another bra-state $\langle L |$,
such as $\langle L | R\rangle$. 
A quantum amplitude also defines a path integral or a partition function $Z$ 
with the linking of worldlines/worldsheets on 
a $d$-manifold $\cM^d$, 
read as
\be \label{eq:qaZ}
\langle L | R\rangle=Z(\cM^d; \text{Link}[\text{worldline, worldsheet}, \dots]).
\ee
%
For example, the $|R\rangle$ state can
represent a ground state of 2-torus $T^2_{xy}$ 
if we put the system on a solid torus $D^2_{xt} \times S^1_y$ \footnote{This is consistent with the fact that TQFT assigns
  a complex number to a closed manifold without boundary, while assigns a
  state-vector in the Hilbert space to an open manifold with a boundary.}
(see Fig.~\ref{lnklp}(a) as the product space of 2-dimensional disk $D^2$ and 1-dimensional circle $S^1$, where the footnote subindices label the coordinates). 
Note that its boundary 
is $\partial (D^2 \times S^1)=T^2$, and we can view the time $t$ evolving along the radial direction.
We label the trivial vacuum sector without any operator insertions 
as $| 0_{D^2_{xt} \times S^1_y} \rangle$, which is trivial
respect to the measurement of any contractible line operator along $S^1_x$.
A worldline operator creates a pair of 
anyon and anti-anyon at its end points, if it forms a closed loop then it can be viewed as creating 
then annihilating a pair of anyons 
in a closed trajectory. 
{Our worldline or worldsheet operator corresponds to
  the Wilson, 't Hooft, or other extended operator of the gauge theory, although throughout our
  work, we consider a more generic quantum description without limiting to gauge
 theory or quantum field theory (QFT).}
  
Inserting a line operator $W^{S^1_y}_{\si}$ in the interior of $D^2_{xt} \times S^1_y$ gives a new state 
\be
W^{S^1_y}_{\si} | 0_{D^2_{xt} \times S^1_y} \rangle \equiv | \si_{D^2_{xt} \times S^1_y} \rangle.
\ee
Here ${\si}$ denotes the anyon 
type\footnote{If there is a gauge theory description, then the
  quasi-excitation type of particle ${\si}$ and string $\mu$ would be labeled
  by the representation for gauge charge and by the conjugacy class of the
  gauge group for gauge flux.} 
along the oriented line, see Fig.~\ref{lnklp}.

Insert all possible line operators of all $\si$ can completely span the ground state sectors for 2+1D topological order.
The gluing of
$
\< 0_{D^2_{} \times S^1_{}} |0_{D^2_{} \times S^1_{}}\>
$
computes the path integral 
$Z(S^2 \times S^1)$.
If we view the $S^1$ as a compact time,
this counts the ground state degeneracy (GSD, i.e., the number of ground states, or equivalently the dimensions of ground-state Hilbert space $\dim(\cH)$) on 
a 2D 
spatial sphere $S^2$ without quasiparticle insertions.
 Follow the relation based on \eq{eq:qaZ} and the above, we see that for a generic spatial manifold $M_{\text{space}}$, we have a relation:   
\be
\text{GSD}_{M_{\text{space}}}=\dim(\cH) = |{Z}({M_{\text{space}} \times S^1})|.
\ee
For a partition function without any insertion (i.e., anyonic excitations associated extended operators), the following computes a 1-dimensional Hilbert space on a spatial $S^2$ in 2+1D:
\be
\< 0_{D^2_{} \times S^1_{}} |0_{D^2_{} \times S^1_{}}\>=Z(S^2 \times S^1)=1.
\ee
Similar relations hold for other dimensions, e.g. 3+1D topological orders on a $S^3$ without quasi-excitation 
yields
\be
\< 0_{D^3 \times S^1} |0_{D^3 \times S^1}\>=Z(S^3 \times S^1)=1
.
\ee

\subsection{Elementary Geometric Topology, Notations and Surgery Formulas}

{Below we list down some useful and elementary surgery formulas, or other math formulas}, 
in order to prepare for the later derivation of more involved but physically more interesting surgery process.

\begin{center}
{ 
\begin{table}[!t]
\hspace{-3.8em}\begin{tabular}{l}
\hline
\hline
{Manifolds}:\\
 \hline
\hline
3-manifolds without boundaries:\\
$S^3, \;\; S^2 \times S^1,\;\; (S^1)^3=T^3$, etc.\\
3-manifolds with boundaries: \\ 
$D^3, \;\; D^2 \times S^1$, etc.\\
4-manifolds without boundaries:\\ 
$S^4, \;\; S^3 \times S^1, \;\; S^2 \times S^2, \;\; S^2 \times (S^1)^2 =S^2 \times T^2,\;\; (S^1)^4=T^4$, $S^3 \times S^1 \# S^2 \times S^2$, 
$S^3 \times S^1 \# S^2 \times S^2 \# S^2 \times S^2$. \\
4-manifolds with boundaries: \\
$D^4, \;\; D^3 \times S^1,  \;\;  D^2 \times S^2,  \;\; D^2 \times (S^1)^2 = D^2 \times T^2 \equiv C^4, \;\; S^4 \smallsetminus D^2 \times T^2 $.\\
Certain 2-manifolds as the boundaries of 3-manifolds:\\
$S^2, \;\; (S^1)^2=T^2$, etc.\\
Certain 3-manifolds as the boundaries of 4-manifolds:\\
$S^3, \;\; S^2 \times S^1, \;\; (S^1)^3=T^3$, etc.\\
\hline
\hline
{Surgery}:\\ 
 \hline
\hline
Cutting and gluing 4-manifolds:\\
$S^4=(D^3 \times S^1) \cup_{S^2 \times S^1} (S^2 \times D^2)=(D^2 \times T^2) \cup_{T^3} ({S^4 \smallsetminus D^2 \times T^2})=D^4 \cup D^4$. \\
$S^3 \times S^1= (D^3 \times S^1) \cup_{S^2 \times S^1} (D^3 \times S^1)= (D^2 \times S^1 \times S^1) \cup_{T^3} ({S^1 \times D^2 \times S^1})=
 (D^2 \times T^2) \cup_{T^3; \cS^{xyz}} (D^2 \times T^2)$.\\
 $S^2 \times S^2= (D^2 \times S^2) \cup_{S^2 \times S^1} (D^2 \times S^2)= ({S^4 \smallsetminus D^2 \times T^2}) \cup_{T^3; \cS^{xyz}} ({S^4 \smallsetminus D^2 \times T^2})$.\\
 $S^2 \times S^1  \times S^1= (D^2 \times T^2) \cup_{T^3} (D^2 \times T^2)$.\\
 $S^3 \times S^1  \# S^2 \times S^2 = (S^4 \smallsetminus  D^2 \times T^2) \cup_{T^3; \cS^{xyz}} (D^2 \times T^2)$.\\
  $S^3 \times S^1  \# S^2 \times S^2  \# S^2 \times S^2 = (S^4 \smallsetminus  D^2 \times T^2) \cup_{T^3} (S^4 \smallsetminus  D^2 \times T^2)$.\\
Cutting and gluing 3-manifolds:\\
%
%
$S^3=(D^2 \times S^1) \cup_{T^2;} (S^1 \times D^2)=(D^2 \times S^1) \cup_{T^2;\cS_{xy}} (D^2 \times S^1) =D^3 \cup D^3$.\\
$S^2 \times S^1 =(D^2 \times S^1) \cup_{T^2} (D^2 \times S^1)=(D^2 \times S^1) \cup_{S^1 \times S^1;(\cT_{xy})^n} (D^2 \times S^1) $.\\
%
%
\hline
\hline
{Mapping Class Group (MCG)}:\\
MCG($T^d$)=SL($d,\Z$).  \;\;\;\;\; MCG$({S^2 \times S^1})=\Z_2 \times \Z_2$.
\\
 \hline
\hline
\end{tabular}
\caption{Manifolds, surgery formula and mapping class group (MCG) that are considered in the limited case of
our study.
The $S^n$ is an $n$-dimensional sphere,
the $D^n$ is an $n$-dimensional disk,
and the $T^n$ is an $n$-dimensional torus.
The connected sum of two $d$-dimensional manifolds $M_1 \# M_2$ follows the definition in \eqn{eq:connect}.
The gluing of two manifolds $M_1 \cup_{B;\varphi} M_2$ along their boundary $B$ via the map $\varphi$ follows \eqn{eq:gluing-B-phi}.
The complement space notation $M_1 \!\smallsetminus\! M_2$ follows \eqn{eq:comple}.
The mapping class group MCG($T^d$)=SL($d,\Z$) is a special linear group with the matrix representation of 
integer $\Z$ entries.
{The mapping class group MCG$({S^2 \times S^1})=\Z_2 \times \Z_2$ have two generators:  
A generator for the first $\Z_2$ is given by a homeomorphism that is a reflection of each of $S^2$ and $S^1$ separately, so it preserves the orientation of $S^2 \times S^1$.  
A generator for the second $\Z_2$ has the form $f(x,y) = (g_y(x),y)$  where $g_y$ is the rotation of 
$S^2$ along a fixed axis by an angle that varies from 0 to $2\pi$ as $y$ goes once around $S^1$.
If we do not restrict our attention to homeomorphisms that preserve orientation, 
then this ``extended mapping class group (extended-MCG)'' has twice elements 
whenever the manifold has an orientation-reversing homeomorphism, as in the examples we considered above.  
Thus for $S^1, S^2$, and $S^3$, the extended-MCG has $\Z_2$, while we have the 
extended-MCG$({S^2 \times S^1})=(\Z_2)^3$.
}
}
\label{table:manifolds}
\end{table}
}
\end{center}
Some of the above results are well-known \cite{thurston1997three, gompf19994}, while others are less familiar, and perhaps also novel to the literature.

\begin{enumerate}[label=\textcolor{blue}{\arabic*.}, ref={\arabic*},leftmargin=*]

\item 
For example, we consider the connected sum of two $d$-dimensional manifolds $M_1$ and $M_2$, denoted as
\bea \label{eq:connect}
M_1 \# M_2
\eea 
which becomes
 a  new manifold formed by deleting a ball $D^d$ inside each manifold and gluing together the resulting boundary spheres $S^{d-1}$.

\item
We write the partition function on spacetime that is formed by disconnected manifolds
$M$ and $N$, which is denoted as 
\bea M\sqcup N,
\eea 
obeying:
\begin{align}
Z(M\sqcup N)=Z(M)Z(N).
\end{align}

\item
For a closed manifold $M$ glued by two pieces of $d$-dimensional manifolds $M_U$ and $M_D$, so that we denote the gluing as
\bea
M=M_U \cup_B M_D
\eea where the $M_U$ and $M_D$ share a common boundary $(d-1)$-dimensional manifold 
\be
 B=\partial M_U = \overline{\partial M_D}. 
\ee
Note that ${\partial M_D}$ and $\overline{\partial M_D}$ are differed by a sign of the orientation.
If $M_U$ and $M_D$ are oriented, then the $M=M_U \cup_B M_D$ can inherit an induced natural orientation,
obeying the chosen orientation of $M_U$ and $M_D$.
This requires an identification of $\partial M_U \simeq \overline{\partial M_D}$ by reversing the inherited orientation, 
as an 
homeomorphism.\footnote{Since we only focus on orientable manifolds in this work, there is no subtle problem.
In this work, we will not be particularly interested in the details of orientation, so let we limit to the case that we do not emphasize
$\overline{\partial M_D}$ or $\partial M_D$.
In future work, we will consider the generalization to the case for non-orientable manifolds.}

\item  
Moreover, we can have an extra mapping $\varphi$ allowed by diffeomorphism when gluing two manifolds. 
The notation for gluing the boundaries via the $\varphi$ is written as
\be
\label{eq:gluing-B-phi}
M_U \cup_{B;\varphi} M_D. 
\ee
It requires that the boundary to be the same,
$
\partial M_U = \overline{\partial M_D}=B
$.
In particular, we will focus on
a $\varphi$ of mapping class group (MCG) of the boundary $B$ in our work. Thus, we can apply any element of $\varphi \in \MCG(B)$.
For $\varphi =1$ as a trivial identity map, we can simply denote it as $\cM_1 \cup_{B} \cM_2=\cM_1 \cup_{B,I} \cM_2.$

\item
We denote the complement space of $d$-dimensional  $\cM_2$ out of $\cM_1$ as
\be \label{eq:comple}
{\cM_1 \- \cM_2}.
\ee  
This means we cut out $\cM_2$ out of $\cM_1$.
For example, to understand the connected sum $M_1 \# M_2$,
we can cut a ball $D^d$ ($D$ for the disk $D^d$, which is the same a $d$-dimensional ball) out of the $M_1$ and $M_2$.
Each of $M_1 \- D^d$ and $M_2\- D^d$ has a boundary of a sphere $S^{d-1}$. We glue the two manifolds $M_1$ and $M_2$
by a cylinder $S^{d-1} \times I^1$ where the $I^1\equiv I$ is a 1 dimensional interval.

\end{enumerate}

{Throughout our article, we consistently use $\sigma$
  to represent the quasi-excitations of anyonic particle label (such as charge, electric charge, magnetic monopole, or representation of the gauge group) whose spacetime trajectory become
  1-dimensional worldlines
  $W^{S^1}_{\si}$ of 1-circle. 
  We use $\mu$ to represent the quasi-excitations of anyonic string (or loop, or gauge flux) label 
  whose spacetime trajectory become
2-dimensional
  worldsheets, e.g. $V_{\mu}^{S^2}$ and $V_{\mu}^{T^2}$, for the surface of
  2-sphere and 2-torus.}

We can also derive some helpful homology group formulas, via Alexander duality and other relations:\footnote{
To understand the following equations,
suppose that $X$ is a $d$-dimensional manifold with boundary ($\partial X$),  
there is a long exact sequence in homology:
$$
\dots H_k(\partial X) \to  H_k(X) \to H_k(X, \partial X) \to H_{k-1}(\partial X) \dots.
$$
The first and the left most arrow is an inclusion. 
The middle arrow is modding out by the image of the inclusion. 
The last arrow is the connecting homomorphism.  
Note that an element in $H_k(X, \partial X)$ 
is represented by a singular chain whose boundary is in $\partial X$.  
The right most arrow assigns to such a chain its boundary.\\
There is a dual exact sequence for cohomology:
$$
\dots H^{m-1}(\partial X) \to H^{m}(X, \partial X) \to H^{m}(X) \to H^{m}(\partial X) \dots.
$$
Note that $ H^{m}(X, \partial X)$ is represented by cocycles that are zero on all chains in $(\partial X)$. 
The right most arrow here becomes the connecting homomorphism.  
The other two arrows are simply an algebraic inclusion (the middle one) or a pull-back. 
Duality imposes that 
\bea
H^{d-k-1}(\partial X) &=& H_{k}(\partial X),\\  
H^{d-k}(X,\partial X) &=& H_{k}(X),\\  
H^{d-k}(X) &=& H_{k}(X, \partial X).
\eea 
If we use this result and Alexander duality, we can deduce the equations:
\eqn{eq:H1D}, \eqn{eq:H2D}, \eqn{eq:H1T}, and \eqn{eq:H2T}.
}
\bea
H_1(D^3 \times S^1,\Z)&=&{H_2}(S^4 \- D^3 \times S^1,\Z)={H_2}(D^2 \times S^2,\Z)=\Z, \label{eq:H1D}\\
H_2(D^3 \times S^1,\Z)&=&{H_1}(S^4 \- D^3 \times S^1,\Z)={H_1}(D^2 \times S^2,\Z)=0. \label{eq:H2D}\\
{H_1}(D^2 \times T^2,\Z) &=& H_2({S^4 \- D^2 \times T^2},\Z)=\Z^2, \label{eq:H1T}\\
{H_2}(D^2 \times T^2,\Z) &=& H_1({S^4 \- D^2 \times T^2},\Z)=\Z. \label{eq:H2T}
\eea
Follow the above definitions, we obtain a set of useful formulas, and 
we summarize in Table  \ref{table:manifolds}.\footnote{We thank conversations with Clifford Taubes and correspondences with Robert Gompf clarifying some of these formulas.
For other details, please see also our upcoming work \cite{WWY-Gompf}.
}

\subsection{Data in 2+1D}

\subsubsection{Quantum Fusion Data in 2+1D}

In 2+1D, we 
consider the worldline operators creating particles.
We define the \emph{fusion data} via fusing worldline operators:
\bea
\label{GW}
W^{S^1_y}_{\si_1} W^{S^1_y}_{\si_2}= \cF^\si_{\si_1\si_2} W^{S^1_y}_{\si}, \;\;\text{and }
G^\al_\si \equiv \<\al|  \si_{D^2_{xt} \times S^1_y} \rangle. \;\;
\eea
Here $G^\al_\si$ is read from the projection to a 
complete basis
$ \<\al|$.
Indeed the $W^{S^1_y}_{\si}$ 
generates all the canonical bases from $|0_{D^2_{xt} \times S^1_y}\>$.
Thus the canonical projection can be 
\be
\<\al|=\<0_{D^2_{xt} \times S^1_y}| (W^{S^1_y}_{\al})^\dagger=\<0_{D^2_{xt} \times S^1_y}| (W^{S^1_y}_{\bar{\al}})
=\< \al_{D^2_{xt} \times S^1_y}| ,
\ee
then we have
\be
G^\al_\si  = \<\al|W^{S^1_y}_{\si} |0_{D^2_{xt} \times S^1_y}\>
 =\< 0_{D^2_{xt} \times S^1_y} | (W^{S^1_y}_{\bar{\al}}) W^{S^1_y}_{\si} |0_{D^2_{xt} \times S^1_y}\>
=Z(S^2 \times S^1; {\bar{\al}},\si)=\del_{\al\si},
\ee
where a pair of particle-antiparticle $\si$ and ${\bar{\si}}$ can fuse to the vacuum.
We derive
\bea \label{eq:Nabc}
&&\cF^\al_{\si_1\si_2} =  \<\al|W^y_{\si_1} W^y_{\si_2}|0_{D^2_{xt} \times S^1_y}\> \nonumber\\
&&=\< 0_{D^2_{xt} \times S^1_y} | (W^{S^1_y}_{\bar{\al}}) W^{S^1_y}_{\si_1} W^{S^1_y}_{\si_2}|0_{D^2_{xt} \times S^1_y}\> \nonumber \\
&&=Z(S^2 \times S^1; {\bar{\al}},\si_1,\si_2) \equiv \cN^\al_{\si_1\si_2},
\eea
\cblue{where this path integral 
counts the dimension of the Hilbert space (namely the GSD or the number of channels
$\si_1$ and $\si_2$ can fuse to $\al$) on the spatial $S^2$.
This shows the fusion data $\cF^\al_{\si_1\si_2}$ is 
equivalent to the fusion rule $\cN^\al_{\si_1\si_2}$, symmetric under exchanging $\si_1$ and ${\si_2}$.\footnote{
For readers who require some more background knowledge about the fusion algebra of anyons, 
Ref.~\cite{freedman2003topological, preskill2004lecture, Kitaevhoneycomb, nayak2008non, pachos2012introduction} provide
for an introduction and a review of 2+1D case. Our present work will also study the 3+1D case.}
} 
%

\subsubsection{Quantum Braiding Data in 2+1D}

More generally we can glue the $T^2_{xy}$-boundary of $D^2_{xt} \times S^1_y$ via its mapping class group (MCG), namely $\MCG(T^2)=\SL(2,\Z)$ of the special linear group, 
(see Table \ref{table:manifolds})
generated by 
\bea
\hat{\cS}=\bpm 0&-1\\1&0\epm,\;\; \; \hat{\cT}=\bpm 1&1\\0&1\epm.
\eea
The $\hat{\cS}$ identifies $(x,y) \to (-y, x)$, 
while $\hat{\cT}$ identifies $(x,y) \to (x+y, y)$ of $T^2_{xy}$.\footnote{We may also denote
the $\hat{\cS}$ as $\hat{\cS}^{xy}$
and the $\hat{\cT}$ as $\hat{\cT}^{xy}$.
} 
Based on Eq.(\ref{eq:qaZ}), we write down
the quantum amplitudes of the two $\SL(2,\Z)$ generators $\hat{\cS}$ and $\hat{\cT}$
projecting to degenerate ground states. 
%
We denote gluing two open-manifolds $\cM_1$ and $\cM_2$ along their boundaries $\cB$ under the MCG-transformation $\hat{\cal U}$ to a new manifold 
as $\cM_1 \cup_{\cB; \hat{\cal U}} \cM_2$.\footnote{{We may simplify the gluing notation $\cM_1
  \cup_{\cB; \cal U} \cM_2$ to $\cM_1 \cup_{\cB} \cM_2$ if the mapping class
  group's generator $\cal U$ is trivial or does not affect the glued manifold.
  We may simplify the gluing notation further to $\cM_1 \cup \cM_2$ if the
  boundary ${\cB}=\partial \cM_1=\overline{\partial \cM_2}$ is obvious or stated in the
  text earlier.}
} 
%
Below we introduce three sets of quantum braiding statistics  data in 2+1D,

\begin{enumerate}[label=\textcolor{blue}{\arabic*.}, ref={\arabic*},leftmargin=*]

\item 
Modular SL(2,$\Z$) modular data $\cS$:
It is amusing 
to visualize
the gluing
\be
D^2_{} \times S^1_{} \cup_{T^2; \hat \cS} D^2_{} \times S^1_{} =S^3
\ee
shows 
that 
the $\cS_{\bar\si_1\si_2}$ 
represents the Hopf link of two $S^1$ worldlines $\si_1$ and $\si_2$ (e.g. Fig.\ref{lnklp}(b)) in $S^3$ 
with the given orientation (in the canonical basis  
$\cS_{\bar\si_1\si_2}=\<{\si_1}_{} | \hat\cS | {\si_2}_{} \>$):
\bea
\label{S12}
\cS_{\bar\si_1\si_2}  &
\equiv \<{\si_1}_{D^2_{xt}\times S^1_y} | \hat\cS | {\si_2}_{D^2_{xt}\times S^1_y} \>
=Z \bpm \includegraphics[scale=0.35]{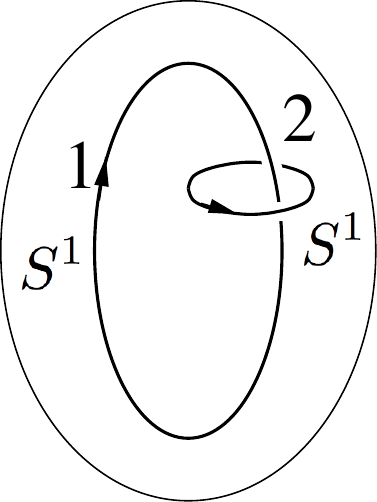} \includegraphics[scale=0.3]{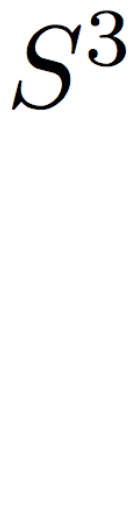} 
 \epm    
.  \;\;\;\;\;
\eea
We note that this data $\cS$ has also been introduced in the work of Witten on TQFT via surgery theory \cite{Witten:1988hf}.
However, here we actually consider more generic quantum mechanical system (a quantum theory, or a quantum many body theory) that does not necessarily requires a QFT description. 
We \emph{only} require the gluing of quantum amplitudes written in the quantum mechanical bra and ket bases living in a Hilbert space associated to the quantum theory 
on spacetime (sub)manifolds.

\item
Modular SL(2,$\Z$) modular data $\cT$:
Use the gluing 
\be 
D^2_{} \times S^1_{} \cup_{T^2; \hat \cT} D^2_{} \times S^1_{}=S^2 \times S^1,
\ee
we can derive a well known result written in the canonical bases, 
\bea
\label{T12}
\cT_{\si_1\si_2} \equiv  \<{\si_1}_{D^2_{xt}\times S^1_y} | \hat\cT | {\si_2}_{D^2_{xt}\times S^1_y} \> 
=\del_{\si_1\si_2}\mathrm{e}^{\ii \th_{\si_2}}.\;\;\;\;\;
\eea
Its spacetime configuration is that two unlinked closed worldlines $\si_1$ and $\si_2$,
with the worldline $\si_2$ twisted by $2\pi$.
The amplitude of a twisted worldline
is given by the amplitude of untwisted worldline multiplied by  
$\mathrm{e}^{\ii \th_{\si_2}}$, where 
$\th_\si/2\pi$ is the spin of the $\si$ excitation.

In summary, the above SL(2,$\Z$) modular data 
implies that $\cS^\text{}_{\bar\si_1\si_2}$ measures
the \emph{mutual braiding statistics} of $\si_1$-and-$\si_2$,
while $\cT_{\si \si}$ measures the \emph{spin} and \emph{self-statistics} of $\si$.

\item
We can introduce additional data, the Borromean rings (BR) linking between three $S^1$ circles in $S^3$, 
written as a path integral $Z$ data with insertions. We denote this path integral $Z$ data as
\bea \label{SBR}
Z \bpm \includegraphics[scale=0.5]{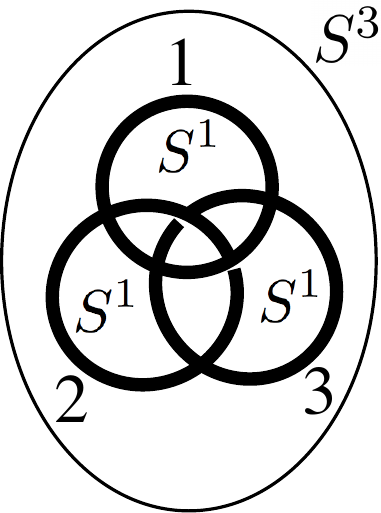} \epm
\cblue{\equiv}Z[S^3; \text{BR}[\sigma_1,\sigma_2,\sigma_3]].
\eea
Although we do not know a bra-ket expression for this amplitude,
we can reduce this configuration to an easier one 
$Z[T^3_{xyt};\sigma'_{1x},\sigma'_{2y},\sigma'_{3t}]$,
a path integral 
of 3-torus $T^3$ 
with three orthogonal line operators each inserting along a non-contractible $S^1$ direction.
The later is a simpler expression because 
we can uniquely 
define the three line insertions exactly along the homology group generators of $T^3$, namely
$H_1(T^3,\Z)=\Z^3$. 
%
The two path integrals 
are related by three consecutive modular SL(2,$\Z$)'s $\cS$ surgeries (or $\cS^{xy}$ surgeries) done along the $T^2$-boundary of $D^2 \times S^1$ tubular neighborhood around three $S^1$ rings. 
{The three-step surgeries we describe earlier
  sequently send the initial 3-sphere configuration with Borromean rings
  insertion 
  \be
  S^3 \xrightarrow{\text{1st surgery}} S^2 \times S^1
  \xrightarrow{\text{2nd surgery}} S^2 \times S^1 \# S^2 \times S^1
  \xrightarrow{\text{3rd surgery}} T^3
  \ee 
  to a 3-torus configuration. Here we use
  the notation $\cM_1 \# \cM_2$ means the connected sum of manifolds $\cM_1$
  and $\cM_2$.}
Namely,
\be
Z[T^3_{xyt};\sigma'_{1x},\sigma'_{2y},\sigma'_{3t}]=
\underset{\small{\sigma_{1},\sigma_{2},\sigma_{3}}}{\sum}
\cS_{ \sigma'_{1x} \sigma_1}
\cS_{ \sigma'_{2y} \sigma_2}
\cS_{ \sigma'_{3z} \sigma_3} Z[S^3; \text{BR}[\sigma_1,\sigma_2,\sigma_3]].
\ee


\end{enumerate}

\subsection{Data in 3+1D}

In {3+1D}, there are intrinsic meanings of braidings of string-like excitations.
We need to consider both the worldline and the worldsheet operators which create particles and strings.
In addition to the $S^1$-worldline operator $W^{S^1}_{\si}$, 
we introduce $S^2$- and $T^2$-worldsheet operators as $V_{\mu}^{S^2}$ and $V_{\mu'}^{T^2}$ which create closed-strings (or loops) at their spatial cross sections. 
We consider the vacuum sector ground state on open 4-manifolds:
\be
\text{
$| 0_{D^3 \times S^1} \rangle $, 
$| 0_{D^2 \times S^2} \rangle$, $| 0_{D^2 \times T^2} \rangle$ and $| 0_{S^4 \- D^2 \times T^2} \rangle$,
}
\ee
while their boundaries are 
\be
\text{$\partial({D^3 \times S^1})=\partial({D^2 \times S^2})=S^2 \times S^1$
and $\partial({D^2 \times T^2})=\partial({S^4 \- D^2 \times T^2})=T^3$.
}
\ee
Here ${\cM_1 \- \cM_2}$ means the complement space of $\cM_2$ out of $\cM_1$. 

\subsubsection{Quantum Fusion Data in 3+1D}

Similar to 2+1D, we define the \emph{fusion data} $\cF^{M}$ by fusing operators: 
\bea
&&W^{S^1}_{\si_1} W^{S^1}_{\si_2}= (\cF^{S^1})^\si_{\si_1\si_2} W^{S^1}_{\si},  \label{F3+1DS1}\\
&&V^{S^2_{}}_{\mu_1} V^{S^2_{}}_{\mu_2}=  (\cF^{S^2})_{{\mu_1}{\mu_2}}^{\mu_3} V^{S^2_{}}_{\mu_3}, \label{F3+1DS2} \\
&&V^{T^2_{}}_{\mu_1} V^{T^2_{}}_{\mu_2}=  (\cF^{T^2})_{{\mu_1}{\mu_2}}^{\mu_3} V^{T^2_{}}_{\mu_3}. \label{F3+1DT2}
\eea
Notice that we introduce additional upper indices in the fusion algebra $\cF^{M}$ to specify the topology of $M$ for 
 the fused operators.\footnote{To be further more specific, we can also specify the
  whole open-manifold topology ${M \times V}$ corresponding to the ground state
  sector $|0_{{M \times V}} \>$, namely we can rewrite the data in new
  notations to introduce the refined data: $(F^{S^1}) \to (F^{D^3 \times
  S^1})$, $(F^{T^2}) \to (F^{D^2 \times T^2})$ and $(F^{S^2}) \to (F^{D^2
  \times S^2})$. However, because $(F^{M})$ are the fusion data in the local
  neighborhood around the worldline/worldsheet operators with topology $M$,
  physically it does not encode the information from the remained product space
  ${M \times V}$. Namely, at least for the most common and known theory that we
  can regularize on the lattice, we understand that $(F^{M})=(F^{M \times
  V_1})=(F^{M \times V_2})=\dots$ for any topology $V_1, V_2, \dots$. So here
  we make a physical assumption that $(F^{S^1}) = (F^{D^3 \times S^1})$,
  $(F^{T^2}) = (F^{D^2 \times T^2})$ and $(F^{S^2}) = (F^{D^2 \times S^2})$.}

We require normalizing worldline/sheet operators for a proper basis, 
so that the $\cF^{M}$ is also properly normalized in order for $Z(  Y^{d-1} \times S^1; \dots )$ as the GSD on
a spatial closed manifold $Y^{d-1}$ always be a positive integer.
In principle, 
we can derive the fusion rule
of excitations in any closed spacetime 4-manifold.
For instance, the fusion rule for fusing three particles on a spatial $S^3$ is
\be
Z(S^3 \times S^1; {\bar{\al}},\si_1,\si_2)
=\langle 0_{D^3 \times S^1} | W^{S^1}_{\bar{\al}}  W^{S^1}_{\sigma_1} W^{S^1}_{\sigma_2} | 0_{D^3 \times S^1} \rangle=
(\cF^{S^1})^\alpha_{\si_1  \sigma_2 }.
\ee
Many more examples of fusion rules can be derived from
computing  
\be
Z(\cM^4; {\si}, {\mu}, \dots) 
\ee
by using $\cF^{M}$ and Eq.(\ref{eq:qaZ}),
here the worldline and worldsheet are submanifolds \emph{parallel not linked} 
 with each other.
 
 {Throughout our work, we consistently use $\sigma$
  to represent the quasi-particle label (such as charge, electric charge, magnetic monopole, or representation of the gauge group) for worldlines
  $W^{S^1}_{\si}$, and we use $\mu$ to represent the quasi-string label for
  worldsheets, e.g. $V_{\mu}^{S^2}$ and $V_{\mu}^{T^2}$.}
 
Overall we choose the operators $W^{S^1}_{\sigma}$ and $V^{M^2}_{\mu}$ carefully, 
so that
they generate linear-independent states when acting on the $|0_{M'} \rangle$ state.

\subsubsection{Quantum Braiding Data in 3+1D}
If the worldline and worldsheet are linked as Eq.(\ref{eq:qaZ}), then the path integral 
encodes 
the \emph{braiding data}. Below we discuss the important braiding processes in 3+1D.
We consider the following four braiding process in 4 dimensional spacetime, thus four sets of quantum braiding data in 3+1D.

\begin{enumerate}[label=\textcolor{blue}{\arabic*.}, ref={\arabic*},leftmargin=*]

\item 
First, the Aharonov-Bohm particle-loop braiding  
can be represented as a $S^1$-worldline of particle and a $S^2$-worldsheet of loop linked in $S^4$ spacetime,
\bea
\label{S2S1glue} 
{\tL}^{(S^2,S^1)}_{ \mu \sigma}
\equiv \langle 0_{D^2 \times S^2} | V_{\mu}^{S^2_{}\dagger} W^{S^1}_{\si} | 0_{D^3 \times S^1} \rangle
=Z \bpm \includegraphics[scale=0.5]{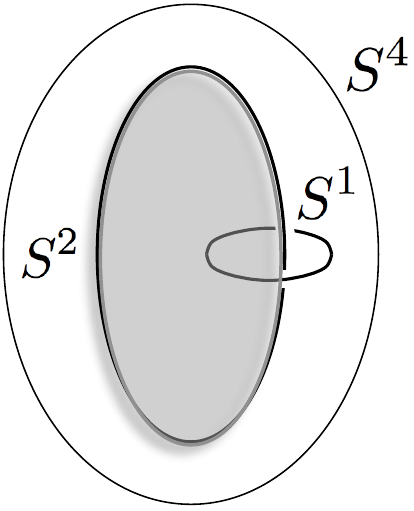} \epm,   
\eea
if we design the worldline and worldsheet along the generators of the first and the second homology group 
$$
H_1(D^3 \times S^1,\Z)={H_2}(D^2 \times S^2,\Z)=\Z
$$ 
respectively, 
via Alexander duality. 
We also use the fact 
\bea 
S^2_{} \times D^2_{}  \cup_{S^2 \times S^1}  D^3_{} \times S^1_{} =S^4,
\eea
thus 
\bea
\langle 0_{D^2 \times S^2 } | 0_{D^3 \times S^1} \rangle= Z(S^4).
\eea

\item
Second, we can also consider 
particle-loop (Aharonov-Bohm) braiding  
as a $S^1$-worldline of particle and a $T^2$-worldsheet 
(below $T^2$ drawn as a $S^2$ with a handle) 
of loop linked in $S^4$,
\bea
\label{T2S1glue}
 \langle  0_{D^2 \times T^2}  | V_{\mu}^{T^2_{}\dagger} W^{S^1}_{\si}   | 0_{S^4 \- D^2 \times T^2} \rangle
=Z \bpm \includegraphics[scale=0.5]{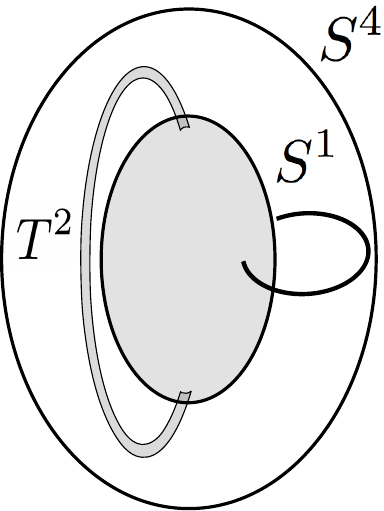} \epm,\;\;\;\;\;
\eea
if we design the worldline and worldsheet along the generators of 
of the first and the second homology group
$$
H_1({S^4 \- D^2 \times T^2},\Z)={H_2}(D^2 \times T^2,\Z)=\Z
$$ 
respectively, 
via Alexander duality.
Compare Eqs.(\ref{S2S1glue}) and (\ref{T2S1glue}),
 the loop excitation of $S^2$-worldsheet 
 is shrinkable, 
while the loop of $T^2$-worldsheet needs not to be shrinkable. 
{If there is a gauge theory description, then the
  loop is shrinkable implies the loop is a pure flux excitation without any net
 charge.}

\item
Third, 
we can 
represent 
a three-loop braiding process \cite{Wang:2014xba, Jiang:2014ksa, Moradi:2014cfa, Wang:2014oya} 
as three $T^2$-worldsheets
 \emph{triple-linking} \cite{carter2004surfaces} in the spacetime $S^4$ (as the first figure in Eq.(\ref{T2T2T2glue})).
We find that
\bea \label{T2T2T2glue}
&&
{\tL^{\text{Tri}}_{{\mu_3}, {\mu_2}, {\mu_1}}} 
\equiv
\langle 0_{S^4 \- {D^2_{wx} \times T^2_{yz} }}| V^{T^2_{zx} \dagger}_{\mu_3} V^{T^2_{xy} \dagger}_{\mu_2} V^{T^2_{yz}}_{\mu_1} | 0_{{D^2_{wx} \times T^2_{yz} }} \rangle \nonumber\\
&&=Z \bpm \includegraphics[scale=0.55]{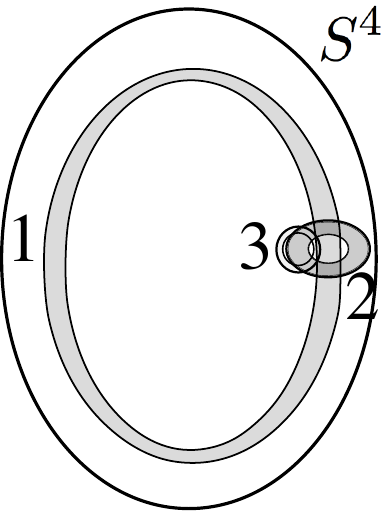} \epm
=Z \bpm \includegraphics[scale=0.55]{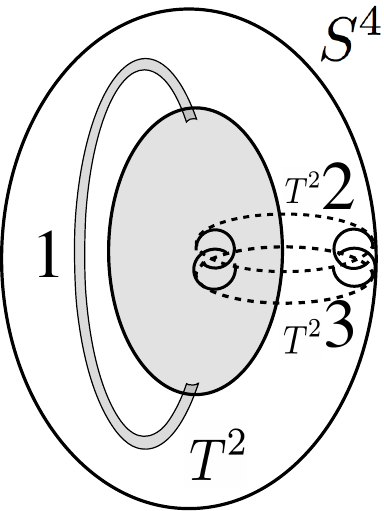} \epm,
\eea
where we design the worldsheets $V^{T^2_{yz}}_{\mu_1}$ along the generator of homology group 
$
H_2(D^2_{wx} \times T^2_{yz},\Z)=\Z
$
while we design  $V^{T^2_{xy} \dagger}_{\mu_2}$ and $V^{T^2_{zx} \dagger}_{\mu_3}$  along the two generators of 
$
{H_2}({S^4\- {D^2_{wx} \times T^2_{yz}}},\Z)=H_1(D^2_{wx} \times T^2_{yz},\Z)=\Z^2
$ 
respectively. 
Again, we obtain via Alexander duality that 
$$
{H_2}({S^4\- {D^2_{} \times T^2_{}}},\Z)=H_1(D^2_{} \times T^2_{},\Z)=\Z^2.
$$
\cblue{We find that Eq.(\ref{T2T2T2glue}) 
is also equivalent to the 
spun surgery construction (or the so-called spinning surgery) 
of a Hopf link (denoted as $\mu_2$ and $\mu_3$) 
linked by a third $T^2$-torus (denoted as $\mu_1$) \cite{Jian:2014vfa, Bi:2014vaa}. 
Namely, we can view the above figure as 
a Hopf link of two loops 
spinning along the dotted path of a $S^1$ circle, which
 becomes a pair of $T^2$-worldsheets $\mu_2$ and $\mu_3$.
Additionally the $T^2$-worldsheet $\mu_1$ 
(drawn in gray as a $S^2$ added a thin handle), 
together with $\mu_2$ and $\mu_3$, the three worldsheets have a 
triple-linking topological invariance \cite{carter2004surfaces}.
}
%

\item Fourth, we consider the four-loop braiding process, 
where three loops dancing in the Borromean ring trajectory while linked by a fourth loop \cite{PhysRevB.91.165119}, 
can characterize certain 3+1D non-Abelian topological orders \cite{Wang:2014oya}. 
We find it is also the spun surgery 
construction of Borromean rings of three loops linked by a fourth torus in the spacetime picture,
and 
its path integral 
$Z[S^4; \text{Link[Spun[BR}[\mu_4 ,\mu_3, \mu_2]],\mu_1]]$ can be transformed:
\bea \label{ZSpinBRS4}
&& 
Z[S^4; \text{Link[Spun[BR}[\mu_4 ,\mu_3, \mu_2]],\mu_1]] \equiv Z \bpm \includegraphics[scale=0.6]{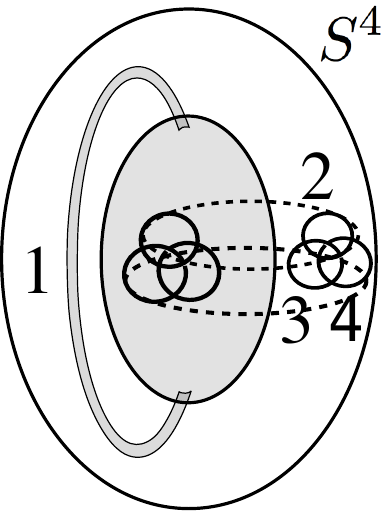} \epm
  \xrightarrow{\;\text{surgery}\;}
Z[T^4 \# S^2  \times S^2; \mu_4',\mu_3',\mu_2',\mu_1' ] \nonumber \\
&&=\langle 0_{T^4 \# S^2 \times S^2 \- {D^2_{wx} \times T^2_{yz}}} |  V^{T^2\dagger}_{\mu_4' } V^{T^2\dagger}_{\mu_3' }  V^{T^2\dagger}_{\mu_2' } V^{T^2_{yz}}_{\mu_1' } |0_{D^2_{wx} \times T^2_{yz}} \rangle,\;\;\;\;\;\;\;\;
\eea
where the surgery 
contains four consecutive modular $\cS$-transformations done along the $T^3$-boundary of $D^2 \times T^2$ tubular neighborhood around four $T^2$-worldsheets. 

{The four-step surgeries here sequently send the
  initial configuration 4-sphere $S^4$ into: 
  \bea
  S^4 &&= (D^3 \times S^1) \cup (S^2 \times
  D^2) = ( (S^3 \- D^3) \times S^1) \cup (S^2 \times D^2)\\
&&  \xrightarrow{\text{1st surgery}} ( (S^2 \times S^1 \- D^3) \times S^1) \cup
  (S^2 \times D^2)  \nn\\
 && \xrightarrow{\text{2nd surgery}} ( (S^2 \times S^1\# S^2
  \times S^1 \- D^3) \times S^1) \cup (S^2 \times D^2) \nn \\
 && \xrightarrow{\text{3rd
  surgery}} ( (T^3 \- D^3) \times S^1) \cup (S^2 \times D^2) = (T^4 \- D^3
  \times S^1) \cup (S^2 \times D^2) \nn \\
 && \xrightarrow{\text{4th surgery}} T^4 \# S^2
  \times S^2 \nn
  \eea
  to a connected sum of $T^4$-torus and $S^2 \times S^2$
  configuration. Here we use the notation $\cM_1 \# \cM_2$ means the connected
  sum of manifolds $\cM_1$ and $\cM_2$. The outcome $Z[T^4 \# S^2 \times S^2;
  \mu_4',\mu_3',\mu_2',\mu_1' ]$ has the wonderful desired property that we can
  design the worldsheets along the generator of homology groups so the operator
  insertions are well-defined. Here $V^{T^2_{yz}}_{\mu_1' }$ is the worldsheet
  operator acting along the only generator of homology group $H_2(D^2 \times
  T^2,\Z)=\Z$. And $V^{T^2}_{\mu_2' }, V^{T^2}_{\mu_3' }, V^{T^2}_{\mu_4' }$
  are the worldsheet operators acting along three among the seven generators of
  $H_2({T^4 \# S^2 \times S^2 \- D^2 \times T^2},\Z)=\Z^7$ with the hollowed
  $D^2 \times T^2$ specified in the earlier text. This shows that 
\bea
Z[T^4 \# S^2
  \times S^2; \mu_4',\mu_3',\mu_2',\mu_1' ]
  \label{eq:ZT4S2}
\eea 
is a better quantum number easier
  to be computed than $Z[S^4; \text{Link[Spun[BR}[\mu_4 ,\mu_3,
  \mu_2]],\mu_1]]$.}
The final spacetime manifold that we obtained after the four-step surgery above is $T^4 \# S^2  \times S^2$, where $\#$ stands for the connected sum.

\item
Modular SL(3,$\Z$) modular data ${\cS}^{xyz}$ and ${\cT}^{xy}$:

We can glue the $T^3$-boundary of 4-submanifolds (e.g. $D^2 \times T^2$ and ${S^4 \- D^2 \times T^2}$) via $\MCG(T^3)=\SL(3,\Z)$ 
generated by  
\bea
\hat{\cS}^{xyz}=\bpm 0& 0&1 \\1& 0&0 \\0& 1&0\epm, \;\;\; \hat{\cT}^{xy}=\bpm 1&1 &0\\0&1 & 0 \\0 &0 & 1\epm.
\eea
In this work, we define their representations as 
\bea
&& \label{eq:Sxyz}
{\cS^{xyz}_{\mu_2, \mu_1}} \equiv 
\< 0_{D^2_{xw} \times T^2_{yz}} | V^{T^2_{yz} \dagger}_{\mu_2}  \hat{\cS}^{xyz}  V^{T^2_{yz}}_{\mu_1}  | 0_{D^2_{xw} \times T^2_{yz}} \rangle, \\
&& \label{eq:Txy}
{\cT^{xy}_{\mu_2, \mu_1}} 
\equiv \< 0_{D^2_{xw} \times T^2_{yz}} | V^{T^2_{yz} \dagger}_{\mu_2}  \hat{\cT}^{xy}  V^{T^2_{yz}}_{\mu_1}  | 0_{D^2_{xw} \times T^2_{yz}} \rangle,
\eea
while $\cS^{xyz}$ is a spun-Hopf link in $S^3 \times S^1$, and ${\cT}^{xy}$ is related to the \emph{topological spin} and \emph{self-statistics} of closed strings \cite{Wang:2014oya}.

{In this work, we project $\hat{\cS}^{xyz}$ and
  $\hat{\cT}^{xy}$ into the bra-ket bases of $| {\mu}_{D^2 \times T^2} \rangle
  \equiv V^{T^2_{yz}}_{\mu} | {0}_{D^2 \times T^2} \rangle$. Our representation
  of 
  $${\cS^{xyz}_{\mu_2, \mu_1}} \equiv \< {\mu_2}_{D^2 \times T^2}|
  \hat{\cS}^{xyz} | {\mu_1}_{D^2 \times T^2} \rangle =\< 0_{D^2_{xw} \times
  T^2_{yz}} | V^{T^2_{yz} \dagger}_{\mu_2} \hat{\cS}^{xyz} V^{T^2_{yz}}_{\mu_1}
  | 0_{D^2_{xw} \times T^2_{yz}} \rangle =Z[{S^3 \times S^1};
  \text{Spun[Hopf}[{\mu_2},{\mu_1}]]]$$ 
  is effectively a path integral of a
  spun Hopf link in the spacetime manifold 
  $$
  D^2_{} \times T^2_{} \cup_{T^3;
  \hat \cS^{xyz}} D^2_{} \times T^2_{} =S^3 \times S^1.
  $$ 
  Our representation of
  $$
  {\cT^{xy}_{\mu_2, \mu_1}} \equiv \< {\mu_2}_{D^2 \times T^2}| \hat{\cT}^{xy}
  | {\mu_1}_{D^2 \times T^2} \rangle =\< 0_{D^2_{xw} \times T^2_{yz}} |
  V^{T^2_{yz} \dagger}_{\mu_2} \hat{\cT}^{xy} V^{T^2_{yz}}_{\mu_1} |
  0_{D^2_{xw} \times T^2_{yz}} \rangle
  $$ 
  is effectively a $Z[S^2 \times S^1
  \times S^1]$-path integral in the spacetime manifold 
  $$D^2_{} \times T^2_{}
  \cup_{T^3; \hat \cT^{xy}} D^2_{} \times T^2_{} =S^2 \times S^1 \times S^1.$$
  In addition, the worldsheet operator $V^{T^2_{yz}}_{\mu}$ effectively
  contains also worldline operators, e.g. $W^{S^1_{y}}_{}$ and $W^{S^1_{z}}_{}$
  along $y$ and $z$ directions. Namely we mean that $
  V^{T^2_{yz}}_{\mu}=W^{S^1_{y}} W^{S^1_{z}} V^{T^2_{yz}}$, so $W^{S^1_{y}}$
  and $W^{S^1_{z}}$ are along the two generators of homology group $H_1({D^2
  \times T^2},\Z)=\Z^2$, while $V^{T^2_{yz}}$ is along the unique one generator
  of homology group $H_2({D^2 \times T^2},\Z)=\Z$. If there is a gauge theory
  description, then we project our $\hat{\cS}^{xyz}$ and $\hat{\cT}^{xy}$ into
  a one-flux (conjugacy class) and two-charge (representation) basis 
  \be
  |
  {\mu_1}, {\si_2}, {\si_3} \rangle \equiv V^{T^2_{yz}}_{\mu_1}
  W^{S^1_{y}}_{\si_2} W^{S^1_{z}}_{\si_3} | {0}_{D^2_{xw} \times T^2_{yz}}
  \rangle.
\ee 
  So our projection here is different from the one-charge
  (representation) and two-flux (conjugacy class) basis $| {\si_1}, {\mu_2},
  {\mu_3}\rangle$ used in Ref.~\cite{Jiang:2014ksa, Moradi:2014cfa,
  Wang:2014oya}. The $| {\si_1}, {\mu_2}, {\mu_3}\rangle$-bases can be obtained
  through 
  \be \label{eq:basis-s1m2m3}
  | {\si_1}, {\mu_2}, {\mu_3}\rangle \equiv W^{S^1_{x}}_{\si_1} V^{T^2_{xy}}_{\mu_2} V^{T^2_{xz}}_{\mu_3} |
  {0}_{S_4 \- D^2_{xw} \times T^2_{yz}} \rangle,
  \ee
  where $W^{S^1_{x}}$ is along
  the generator of homology group $H_1({S^4 \- D^2 \times T^2},\Z)=\Z$, while
  $V^{T^2_{xy}}_{\mu_2}$ and $V^{T^2_{xz}}_{\mu_3}$ are along the two
  generators of homology group $H_2({S^4 \- D^2 \times T^2},\Z)=\Z^2$ via the
  Alexander duality. The alternate representation of the SL(3,$\Z$) modular data via the bases \eq{eq:basis-s1m2m3}
  is presented in our upcoming work \cite{WWY-Gompf}.
  }
\end{enumerate}

\section{New Quantum Surgery Formulas: Generalized Analogs of Verlinde's}

\subsection{Derivations of some basics of quantum surgery formulas}

Now we like to derive two powerful identities (\eqn{ZMN}, and \eqn{eq:surger2}) for fixed-point path integrals or partition functions
for quantum theory, for example suitable for studying topological orders. 

\begin{enumerate}[label=\textcolor{blue}{\arabic*.}, ref={\arabic*},leftmargin=*]

\item 
If the path integral 
formed by disconnected manifolds
$M$ and $N$, denoted as $M\sqcup N$,
we have
$
Z(M\sqcup N)=Z(M)Z(N)
$.
Assume that\\
(1) we divide both $M$ and $N$ into two pieces such that
$M=M_U\cup_{B} M_D$, $N=N_U\cup_{B} N_D$,
and their cut 
topology (dashed boundary denoted as $B$) 
is equivalent $B=\overline{\prt M_D}={\prt M_U}=\overline{\prt N_D}={\prt N_U}$,
and\\ 
(2) the Hilbert space  on the spatial slice is 1-dimensional  
(namely the GSD=1),\footnote{{Presumably there may be defect-like excitation of
  particles and strings on the spatial slice cross-section $B$. If the
  dimensional of Hilbert space on the spatial slice $B$ is 1, namely the ground
  state degeneracy (GSD) is 1, then we can derive the gluing identity 
\be
\< M_U |
  M_D \>=\< N_U | N_D \> \Rightarrow \< M_U | N_D \>=\< N_U | M_D \> 
\ee 
because
  this vector space (of the Hilbert space) is 1-dimensional and all vectors are parallel in the inner
  product.}
}
then we obtain
\bea
\label{ZMN}
&& Z \bpm \includegraphics[scale=0.33]{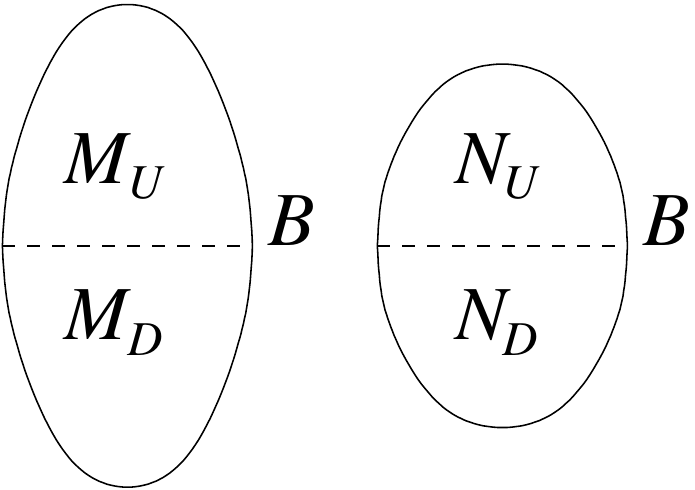} \epm 
=
 Z \bpm \includegraphics[scale=0.33]{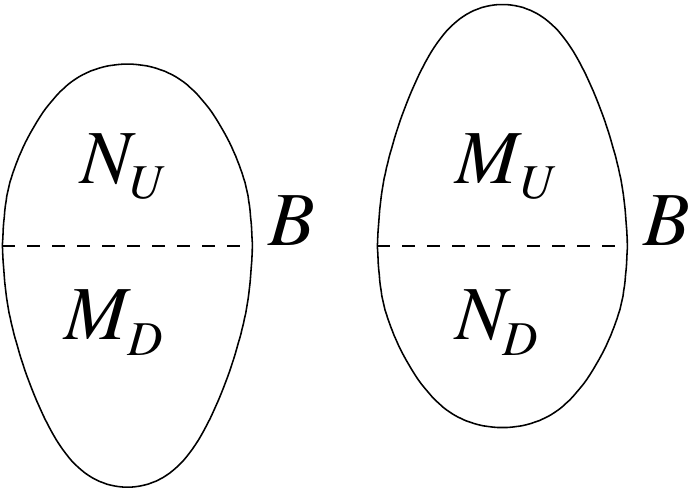} \epm 
\\
\Rightarrow
&&\boxed{Z(M_U \cup_B M_D) 
Z(N_U \cup_B N_D) 
 =
Z( N_U\cup_B  M_D) 
Z(M_U \cup_B N_D )}. \nonumber
\eea

\item

Now we derive a generic formula for our use of surgery. We consider a closed manifold $M$ glued by two pieces $M_U$ and $M_D$ so that
$M=M_U \cup_B M_D$ where $B=\partial M_U =  \overline{\partial M_D}$.
We consider there are insertions of operators in $M_U$ and $M_D$.
We denote the generic insertions in $M_U$ as $\alpha_{M_U}$ and 
the generic insertions in $M_D$ as $\beta_{M_D}$. Here both
$\alpha_{M_U}$ and $\beta_{M_D}$ may contain both worldline and worldsheet operators.
We write the path integral 
as 
$
{Z(M; \alpha_{M_U}, \beta_{M_D})}=\<  \alpha_{M_U} | \beta_{M_D}\>,
$
while the worldline/worldsheet may be linked or may not be linked in $M$.
Here we introduce an extra subscript $M$ in 
$${Z(M; \alpha_{M_U}, \beta_{M_D})}=\<  \alpha_{M_U} | \beta_{M_D}\>_M$$ to specify the glued manifold is $M_U \cup_B M_D=M$. 
Now we like to do surgery by cutting out the submanifold $M_D$ out of $M$ and re-glue it back to $M_U$ via its mapping class group (MCG) generator
\be
\hat{K} \in \MCG(B)= \MCG(\partial M_U) =   \MCG(\overline{\partial M_D}).
\ee 
We now give some additional assumptions.\\
{\bf Assumption 1:} The operator insertions in $M$ are well-separated into $M_U$ and $M_D$, so that no operator insertions cross the
boundary $B$.
Namely, at the boundary cut $B$ there are no defects of point or string excitations from the
cross-section of $\alpha_{M_U}$, $\beta_{M_D}$ or any other operators.\footnote{
The readers should notice that this assumption is stronger and more restricted than
the previous \eqn{ZMN}. In \eqn{ZMN}, we can have 
defects of point or string excitations from the
cross-section and on the boundary cut $B$, as long as the dimension of Hilbert space associated to $B$ is 1-dimensional.}
\\
\noindent
{\bf  Assumption 2:} We can generate the complete bases of degenerate ground states
fully spanning the dimension of Hilbert space for the spatial section of $B$, by inserting distinct operators (worldline/worldsheet, etc.) into $M_D$.
Namely, we insert a set of operators $\Phi$ in the interior of $| 0_{M_D}\> $ to obtain a new state $\Phi | 0_{M_D}\> \equiv | \Phi_{M_D}\>$, 
such that these states $\{\Phi | 0_{M_D}\> \}$ are orthonormal canonical bases, and
the dimension of the vector space $\dim(\{\Phi | 0_{M_D}\> \})$ equals to the ground state degeneracy ($\GSD$) of the topological order on the spatial section $B$.

If both assumptions hold, then we find a relation:
\bea \label{}
&&{Z(M; \alpha_{M_U}, \beta_{M_D})}=\<  \alpha_{M_U} | \beta_{M_D}\>_M  \nonumber\\
&&=\sum_{\Phi} \langle \alpha_{M_U} | \hat{K}  \Phi | 0_{M_D}\> \< 0_{M_D}| (\hat{K}  \Phi)^\dagger |\beta_{M_D} \rangle  \nonumber \\
&&=\sum_{\Phi} \langle \alpha_{M_U} | \hat{K}  \Phi | 0_{M_D}\> \< 0_{M_D}| \Phi^\dagger \hat{K}^{-1} |\beta_{M_D} \rangle  \nonumber \\
&&=\sum_{\Phi} \langle \alpha_{M_U} | \hat{K}   | \Phi _{M_D}\>_{ M_U \cup_{B; \hat{K}} M_D} 
\< \Phi_{M_D}|  \hat{K}^{-1} |\beta_{M_D} \rangle_{ M_D \cup_{B; \hat{K}^{-1}} M_D}  \nonumber \\
&&=\sum_{\Phi} {Z({ M_U \cup_{B; \hat{K}} M_D}; \alpha_{M_U}, \Phi_{M_D})} \< \Phi_{M_D}|  \hat{K}^{-1} |\beta_{M_D} \rangle_{ M_D \cup_{B; \hat{K}^{-1}} M_D} \nonumber\\
&&=\sum_{\Phi} {K}^{-1}_{\Phi, \beta} \; {Z({ M_U \cup_{B; \hat{K}} M_D}; \alpha_{M_U}, \Phi_{M_D})}
\eea
We note that in the second equality, 
we write the identity matrix as $\mathbb{I}=\sum_{\Phi} ( \hat{K}  \Phi ) | 0_{M_D}\> \< 0_{M_D}| (\hat{K}  \Phi)^\dagger$.
In the third and fourth equalities, 
 we have $\hat{K}^{-1}$ in the inner product $ \< \Phi_{M_D}|\hat{K}^{-1} |\beta_{M_D} \rangle$,
because $\hat{K}$ as a MCG generator acts on the spatial manifold $B$ directly.  
The evolution process from the first $\hat{K}^{-1}$ on the right and the second $\hat{K}$ on the left can be viewed as the \emph{adiabatic evolution} of
quantum states in the case of \emph{fixed-point} topological orders.
In the fifth equality, 
we rewrite 
$\langle \alpha_{M_U} | \hat{K}   | \Phi _{M_D}\>_{ M_U \cup_{B; \hat{K}} M_D}={Z({ M_U \cup_{B; \hat{K}} M_D}; \alpha_{M_U}, \Phi_{M_D})}$
where $\alpha_{M_U}$ and $\Phi_{M_D}$ may or may not be linked in the new manifold ${ M_U \cup_{B; \hat{K}} M_D}$.
In the sixth equality, 
we assume that both $|\beta_{M_D} \rangle$ and $|\Phi_{M_D} \rangle$ are vectors in a canonical basis, then
we can define 
\be \label{eq:projectK}
\< \Phi_{M_D}|  \hat{K}^{-1} |\beta_{M_D} \rangle_{ M_D \cup_{B; \hat{K}^{-1}} M_D} \equiv {K}^{-1}_{\Phi, \beta}
\ee 
as a matrix element of ${K}^{-1}$, which now becomes a representation of MCG in the quasi-excitation bases of $\{|\beta_{M_D} \rangle, |\Phi_{M_D} \rangle, \dots \}$.
It is important to remember that ${K}^{-1}_{\Phi, \beta}$ is a quantum amplitude computed in the specific spacetime manifold ${ M_D \cup_{B; \hat{K}^{-1}} M_D}$.

To summarize, so far we derive,
\be \label{eq:surger1}
\boxed{ {Z(M; \alpha_{M_U}, \beta_{M_D})}=\sum_{\Phi} {K}^{-1}_{\Phi, \beta} \; {Z({ M_U \cup_{B; \hat{K}} M_D}; \alpha_{M_U}, \Phi_{M_D})}}.  
\ee
The detailed derivation shows
\bea
&&\sum_{\Phi'} {K}^{}_{\beta, \Phi'}  {Z(M; \alpha_{M_U}, \beta_{M_D})} \nonumber\\
&&=
\sum_{\Phi'} {K}^{}_{\beta, \Phi'} \sum_{\Phi} {K}^{-1}_{\Phi, \beta} \; {Z({ M_U \cup_{B; \hat{K}} M_D}; \alpha_{M_U}, \Phi_{M_D})}  \;\;\;\nonumber\\
&&=
\delta_{\Phi\Phi'}  \; {Z({ M_U \cup_{B; \hat{K}} M_D}; \alpha_{M_U}, \Phi_{M_D})}\nonumber\\
&&={Z({ M_U \cup_{B; \hat{K}} M_D}; \alpha_{M_U}, \Phi'_{M_D})}.  
\eea
We can also derive another formula by applying the inverse transformation,
\be \label{eq:surger2}
\boxed{ {Z({ M_U \cup_{B; \hat{K}} M_D}; \alpha_{M_U}, \Phi'_{M_D})} = \sum_{\Phi'} {K}^{}_{\beta, \Phi'} \; {Z(M; \alpha_{M_U}, \beta_{M_D})}}. 
\ee
if it satisfies ${K} {K}^{-1}=\mathbb{I}$. Again we stress that
${K}^{}_{\beta, \Phi'} $
is a quantum amplitude computed in the specific spacetime manifold ${ M_D \cup_{B; \hat{K}^{-1}} M_D}$.

\end{enumerate}

\subsection{Quantum Surgery Formulas in 2+1D and for 3-manifolds}

In 2+1D, we can derive the 
renowned Verlinde formula \cite{Witten:1988hf,  Verlinde:1988sn, Moore:1988qv}
by one specific version of our \eqn{ZMN}: 
\bea
\label{Z2Dcut}
&&Z \bpm \includegraphics[scale=0.5]{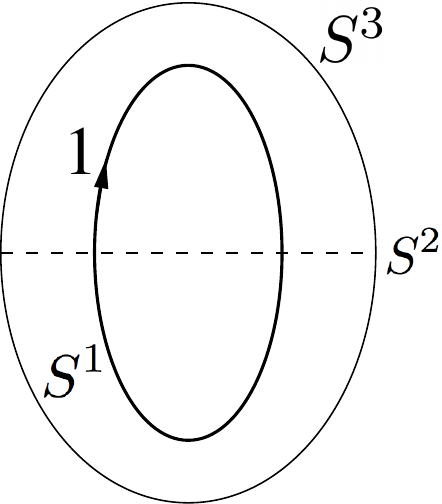} \epm 
 Z \bpm \includegraphics[scale=0.5]{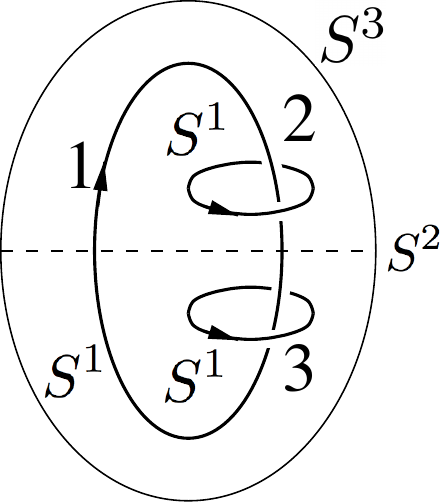} \epm 
=
 Z \bpm \includegraphics[scale=0.5]{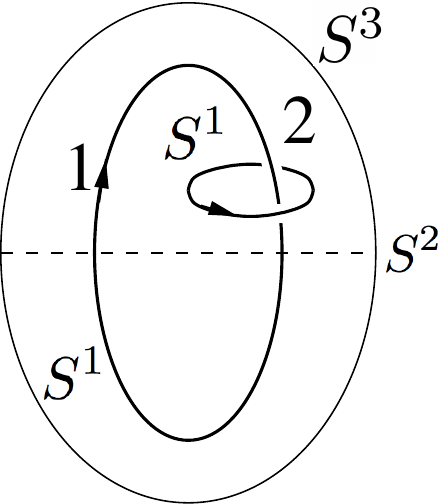} \epm 
 Z \bpm \includegraphics[scale=0.5]{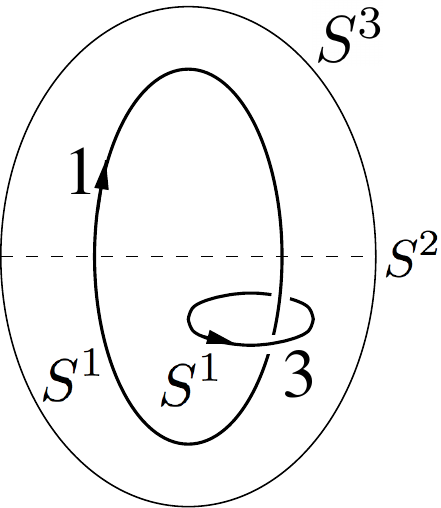} \epm \nonumber\\
&& \quad\quad\quad\quad \quad\quad\quad\quad \;
\Rightarrow  \boxed{\cS^\text{}_{\bar\si_10}\sum_{\si_4} \cS^\text{}_{\bar\si_1\si_4}   \cN^{\si_4}_{\si_2\si_3} 
=
\cS^\text{}_{\bar\si_1\si_2}
\cS^\text{}_{\bar\si_1\si_3}}, 
\eea
where each spacetime manifold is $S^3$, with the line operator insertions such as an unlink and Hopf links.
Each $S^3$ is cut into two $D^3$ pieces, 
so $D^3 \cup_{S^2} D^3={S^3}$, while the boundary dashed cut is $B=S^2$.
The GSD for this spatial section $S^2$ with a pair of particle-antiparticle must be 1, so our surgery satisfies the assumptions for Eq.(\ref{ZMN}).
The second line is derived from rewriting path integrals 
in terms of our data introduced before -- the fusion rule $\cN^{\si_4}_{\si_2\si_3}$ comes from 
fusing ${\si_2\si_3}$ into ${\si_4}$ which Hopf-linked with ${\si_1}$, while Hopf links render the SL(2,$\Z$) modular $\cS$ matrices. 
\cblue{
The label $0$, in $\cS^\text{}_{\bar\si_10}$ and hereafter, denotes a vacuum sector without operator insertions in a submanifold.
}

For Eq.(\ref{Z2Dcut}), the only path integral 
we need to compute more explicitly is this:
\begin{align}
& Z \bpm \includegraphics[scale=0.3]{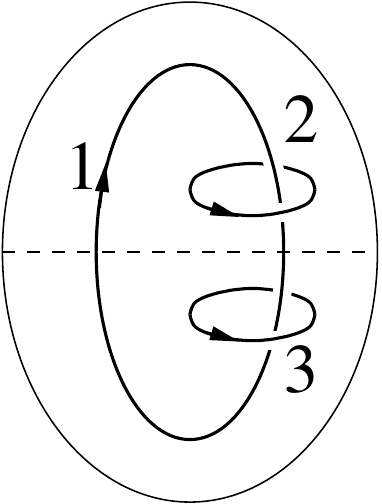} \epm 
= 
 \<0_{D^2_{xt} \times S^1_y}|(W^{S^1_y}_{\si_1})^\dag \hat{\cS} W^{S^1_y}_{\si_2}W^{S^1_y}_{\si_3}|0_{D^2_{xt} \times S^1_y}\> 
\nonumber\\
&= \<0_{D^2_{xt} \times S^1_y}|(W^{S^1_y}_{\si_1})^\dag \hat{\cS} W^{S^1_y}_{\si_4} \cF^{\sigma_4}_{\si_2 \si_3} |0_{D^2_{xt} \times S^1_y}\> 
\nonumber\\
&=
\sum_{\al \si_4} (G^\al_{\si_1})^*\cS_{\al \si_4} \cF^{\si_4}_{\si_2\si_3}=\sum_{\si_4}  \cS_{\bar{\si}_1 \si_4} \cN^{\si_4}_{\si_2\si_3},
\end{align}
where the last equality we use the canonical basis.
Together with the previous data, we can easily derive Eq.(\ref{Z2Dcut}). 

Since it is convenient to express in terms of canonical bases, below for all the derivations, we will implicitly 
project every quantum amplitude into \emph{canonical bases} when we write down its matrix element.

{In the canonical basis when $\cS$ is invertible, we
  can massage our formula to a familiar form, which we derive that:
 \be
 \boxed{
 \cN^{a}_{\si_2\si_3} = \sum_{\bar\si_1}
  \frac{\cS^\text{}_{\bar\si_1\si_2} \cS^\text{}_{\bar\si_1\si_3}
  (\cS^{-1})_{\bar\si_1 a}} {\cS^\text{}_{\bar\si_10}}}.
\ee
  }


\subsection{Quantum Surgery Formulas in 3+1D and for 4-manifolds}

\subsubsection{Formulas for 3+1D particle-string braiding process: Link between 1-worldline and 2-worldsheet}

In 3+1D, we derive that the particle-string braiding process in terms of $S^4$-spacetime path integral Eq.(\ref{S2S1glue}) has the following constraint formulas: 
\bea
\label{eq:S2S1S1inS4}
&& Z \bpm \includegraphics[scale=0.65]{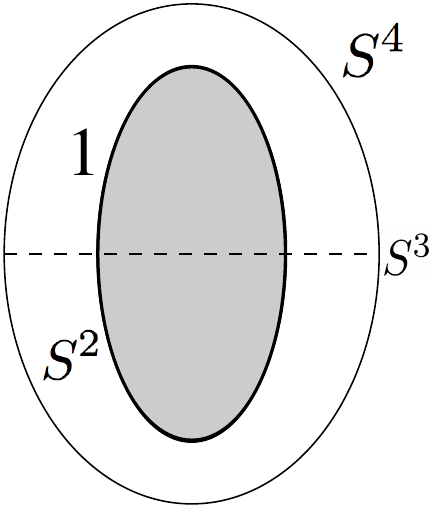}\epm  
  Z \bpm \includegraphics[scale=0.65]{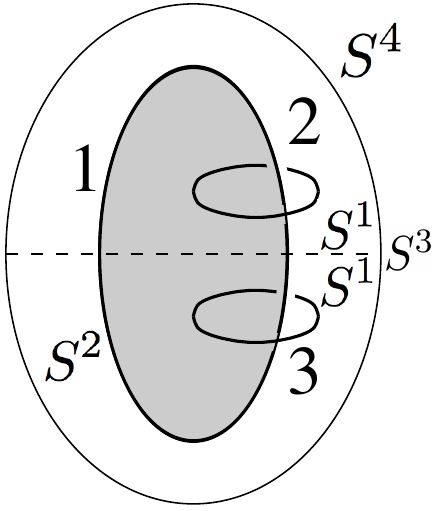} \epm 
=
 Z \bpm \includegraphics[scale=0.65]{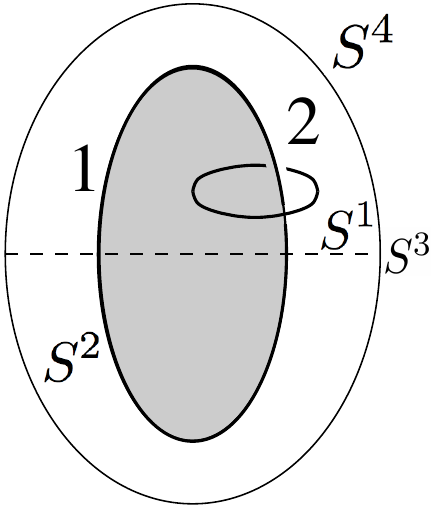} \epm 
 Z \bpm \includegraphics[scale=0.65]{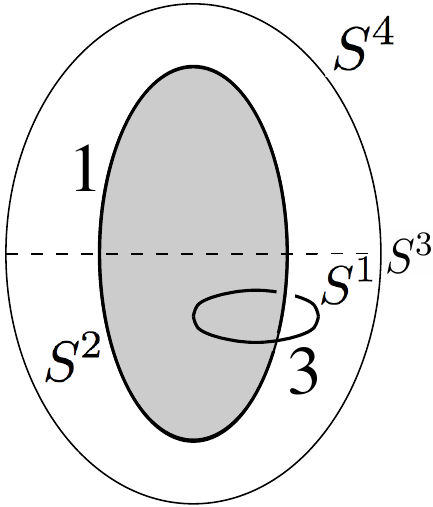} \epm 
\nn \\
&&\quad\quad\quad\quad \quad\quad\quad \; \Rightarrow
\boxed{{{\tL}^\text{($S^2$,$S^1$)}_{\mu_1 0}
\sum_{\si_4}  {\tL}^\text{($S^2$,$S^1$)}_{\mu_1\si_4}  (\cF^{S^1})^{\si_4}_{\si_2\si_3}}
=
{
{\tL}^\text{($S^2$,$S^1$)}_{\mu_1\si_2}
{\tL}^\text{($S^2$,$S^1$)}_{\mu_1\si_3}}}.
 \\
\label{eq:S1S2S2inS4}
 && Z \bpm \includegraphics[scale=0.65]{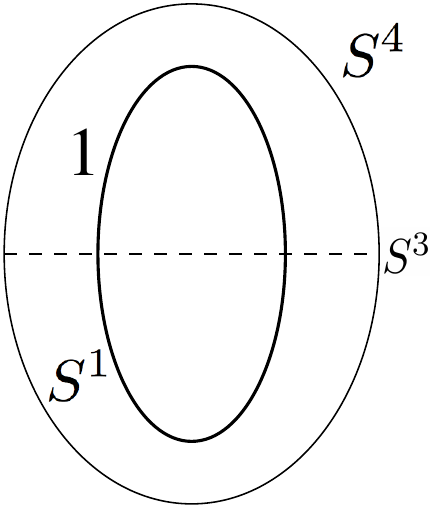} \epm 
 Z \bpm \includegraphics[scale=0.65]{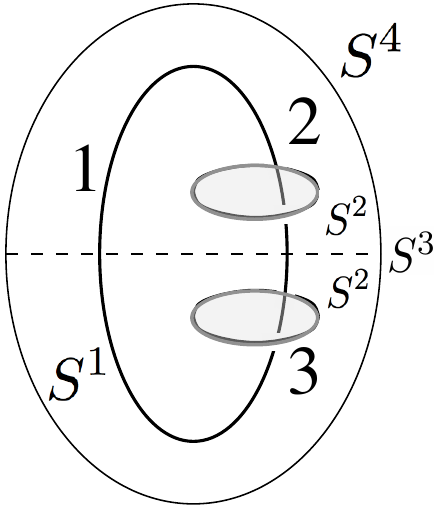} \epm 
=
 Z \bpm \includegraphics[scale=0.65]{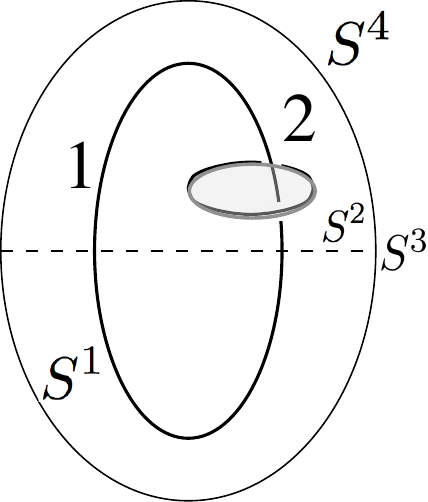} \epm 
 Z \bpm \includegraphics[scale=0.65]{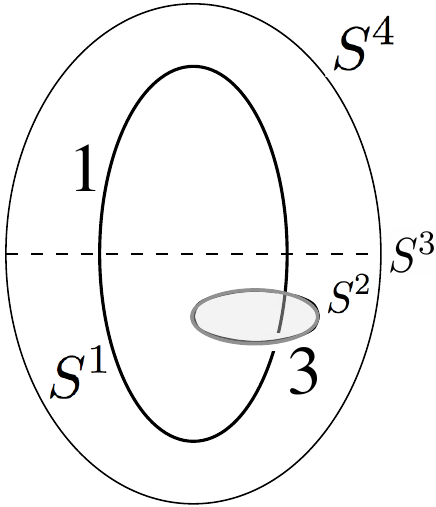} \epm 
\nn \\ 
&&\quad\quad\quad\quad \quad\quad\quad  \; \Rightarrow
\boxed{ 
{{\tL}^\text{($S^2$,$S^1$)}_{0 \si_1}
\sum_{\mu_4}  {\tL}^\text{($S^2$,$S^1$)}_{\mu_4 \si_1}  
(\cF^{S^2})^{\mu_4}_{\mu_2\mu_3}}
=
{
{\tL}^\text{($S^2$,$S^1$)}_{\mu_2 \si_1}
{\tL}^\text{($S^2$,$S^1$)}_{\mu_3 \si_1}}}. \;\;\;\;\;\;\;\;
 \eea
\cblue{Here the gray areas mean $S^2$-spheres.
All the data are well-defined and introduced earlier in Eqs.(\ref{F3+1DS1}), (\ref{F3+1DS2}), and (\ref{S2S1glue})}. 
Notice that Eqs.(\ref{eq:S2S1S1inS4}) and (\ref{eq:S1S2S2inS4}) are symmetric by exchanging worldsheet/worldline indices: $\mu \leftrightarrow \sigma$,
except that the fusion data is different: $\cF^{S^1}$ fuses worldlines, while $\cF^{S^2}$ fuses worldsheets.

For Eq.(\ref{eq:S2S1S1inS4}), the only path integral 
we need to compute more explicitly is this:
\begin{align}
& Z \bpm \includegraphics[scale=0.45]{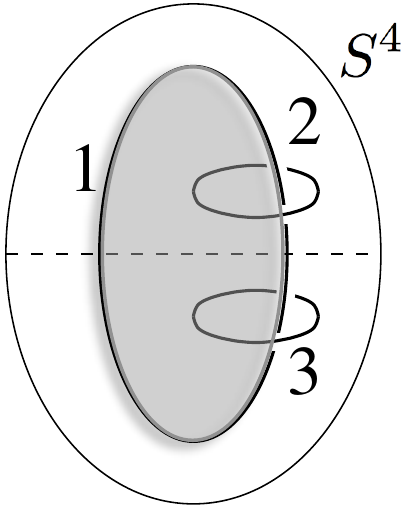} \epm 
= 
 \<0_{D^2_{\varphi w} \times S^2_{\theta\phi}}|(V^{S^2_{\theta\phi}}_{\mu_1})^\dag  W^{S^1_\varphi}_{\sigma_2} W^{S^1_\varphi}_{\sigma_3} | 0_{D^3_{\theta\phi w} \times S^1_{\varphi}} \rangle 
\nonumber\\
&=
 \<0_{D^2_{\varphi w} \times S^2_{\theta\phi}}|(V^{S^2_{\theta\phi}}_{\mu_1})^\dag  W^{S^1_\varphi}_{\sigma_4}  (\cF^{S^1})^{\sigma_4}_{\si_2 \si_3} | 0_{D^3_{\theta\phi w} \times S^1_{\varphi}} \rangle \nonumber\\
&=\sum_{\si_4}  {\tL}^\text{($S^2$,$S^1$)}_{\mu_1\si_4}  (\cF^{S^1})^{\si_4}_{\si_2\si_3},
\end{align}
again we use the canonical basis.
Together with the previous data, we can easily derive Eq.(\ref{eq:S2S1S1inS4}). 
Similarly, we can also derive Eq.(\ref{eq:S1S2S2inS4}), using the almost equivalent computation
following Eq.(\ref{eq:S2S1S1inS4}).


\subsubsection{Formulas for 3+1D three-string braiding process: Triple link between three sets of 2-worldsheets}

We also derive a quantum surgery constraint formula 
for the three-loop braiding process in 3+1D 
in terms of an $S^4$-spacetime path integral Eq.(\ref{T2T2T2glue}) via the ${\cS}^{xyz}$-surgery and its matrix representation:
%
\bea 
\label{eq:Spin[HopfLink]S4}
&&\ \ \ \
  Z \bpm \includegraphics[scale=0.65]{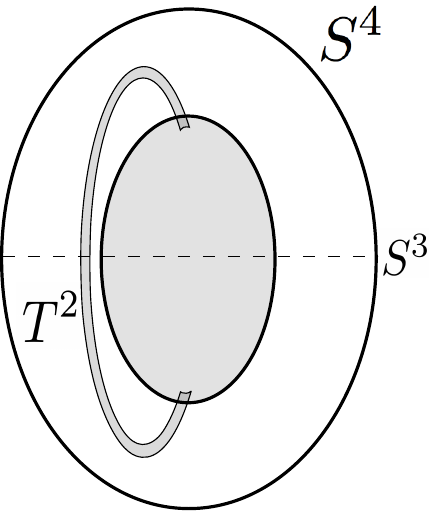} \epm  
 Z \bpm \includegraphics[scale=0.65]{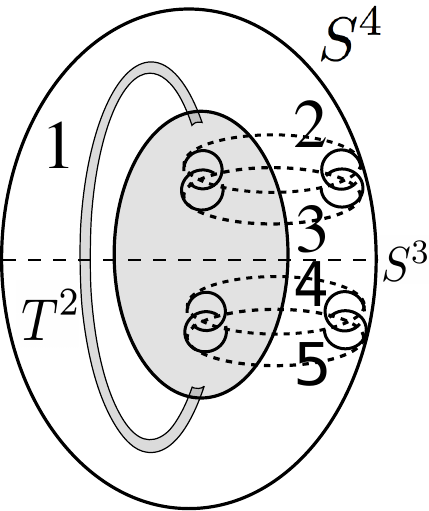} \epm  
_1=
 Z \bpm \includegraphics[scale=0.65]{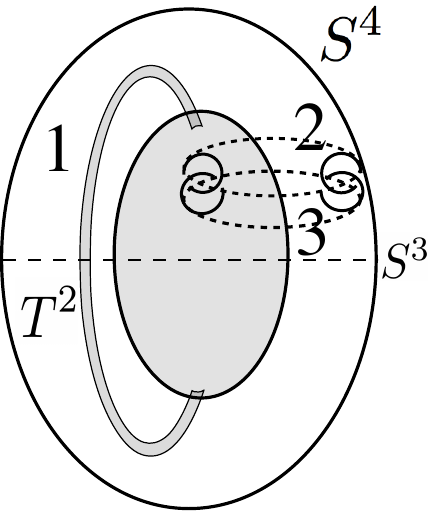} \epm 
 Z \bpm \includegraphics[scale=0.65]{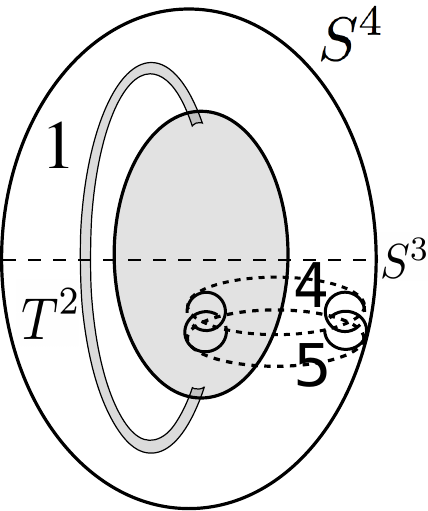} \epm \nonumber \\
&& \Rightarrow
 \boxed{{{\tL^{\text{Tri}}_{0, 0, {\mu_1}}} \cdot
 \sum_{{\Gamma, \Gamma'},{\Gamma_1, \Gamma_1'}}
 (\cF^{T^2})_{{\zeta_2},{\zeta_4}}^{\Gamma}
{(\cS^{xyz})^{-1}_{\Gamma', \Gamma}}  (\cF^{T^2})_{{\mu_1} {\Gamma'}}^{\Gamma_1}  {\cS^{xyz}_{\Gamma_1', \Gamma_1}}  \; {\tL^{\text{Tri}}_{0, 0, {\Gamma_1'}}} }} \nonumber
\\
&&
\boxed{=
{\sum_{{\zeta_2'}, {\eta_2}, {\eta_2'}} 
{(\cS^{xyz})^{-1}_{\zeta_2', {\zeta_2}}}  (\cF^{T^2})_{{\mu_1} {\zeta_2'}}^{\eta_2}  {\cS^{xyz}_{\eta_2', \eta_2}}  \; {\tL^{\text{Tri}}_{0, 0, {\eta_2'}}} } 
\cdot
{\sum_{{\zeta_4'}, {\eta_4}, {\eta_4'}}  
{(\cS^{xyz})^{-1}_{\zeta_4', {\zeta_4}}}  (\cF^{T^2})_{{\mu_1} {\zeta_4'}}^{\eta_4}  {\cS^{xyz}_{\eta_4', \eta_4}}  \; {\tL^{\text{Tri}}_{0, 0, {\eta_4'}}} }}, 
\eea
%
\cblue{here the ${\mu_1}$-worldsheet in gray represents a $T^2$ torus,
while ${\mu_2}$-${\mu_3}$-worldsheets and ${\mu_4}$-${\mu_5}$-worldsheets are both a pair of
two $T^2$ tori obtained by spinning the Hopf link.}
\cblue{All our data 
are well-defined in Eqs.(\ref{F3+1DT2}), (\ref{T2T2T2glue}), and (\ref{eq:Sxyz}) introduced earlier.
For example, the ${\tL^{\text{Tri}}_{0, 0, {\mu_1}}}$ is defined in Eq.(\ref{T2T2T2glue})
with 0 as a vacuum without insertion, so ${\tL^{\text{Tri}}_{0, 0, {\mu_1}}}$ is a
path integral of a $T^2$ worldsheet ${\mu_1}$ in $S^4$.
}
\cblue{The index ${\zeta_2}$ is obtained from} fusing ${\mu_2}$-${\mu_3}$-worldsheets,
and \cblue{${\zeta_4}$  is obtained from} fusing ${\mu_4}$-${\mu_5}$-worldsheets.
Only ${\mu_1},{\zeta_2},{\zeta_4}$ are the fixed indices, other indices are summed over.

Now let us derive Eq.(\ref{eq:Spin[HopfLink]S4}). 
In the first path integral, 
we create a pair of loop $\mu_1$ and anti-loop $\bar{\mu}_1$ excitations and then annihilate them, in terms of
the spacetime picture, we obtain that
\be \label{eq:S4T2}
Z \bpm \includegraphics[scale=0.45]{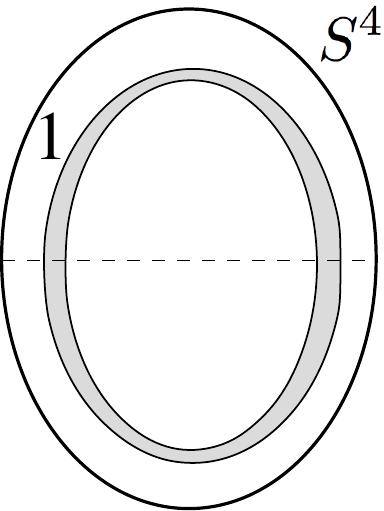} \epm=Z \bpm \includegraphics[scale=0.45]{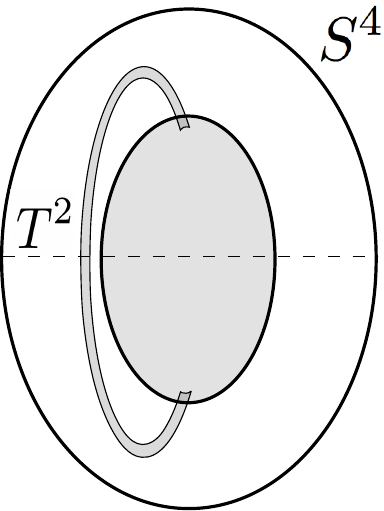} \epm={ \tL^{\text{Tri}}_{0,0,\mu_1}},
\ee
based on the data defined earlier. Let us explain our figure expressions further: \\
--- The grey area drawn in terms of a tube means a 2-torus $T^2$ in topology (the 2-surface insertion in the left hand side of path integral \eqn{eq:S4T2}).\\
--- A 2-torus $T^2$ can be also regarded as a 2-sphere $S^2$ adding a handle in topology (the 2-surface insertion in the right hand side of path integral \eqn{eq:S4T2}).\\

In the third path integral 
${\tL^{\text{Tri}}_{{\mu_3}, {\mu_2}, {\mu_1}}}$, which is Eq.~(\ref{T2T2T2glue}), appeared in Eq.~(\ref{eq:Spin[HopfLink]S4}), there are two descriptions to interpret it in terms of the braiding process in spacetime:
\begin{itemize}
\item
Here is the first description. we create a pair of loop $\mu_1$ and anti-loop $\bar{\mu}_1$ excitations and then
there a pair of  $\mu_2$-$\bar{\mu}_2$ and another pair of $\mu_3$-$\bar{\mu}_3$ are created
while both pairs are thread by $\mu_1$. 
Then the $\mu_1$-$\mu_2$-${\mu_3}$ will do the three-loop braiding process, which gives the most important Berry phase or Berry matrix information into
the path integral. 
After then the pair of  $\mu_2$-$\bar{\mu}_2$ is annihilated and also the pair of $\mu_3$-$\bar{\mu}_3$ is annihilated,
while all the four loops are threaded by $\mu_1$ during the process.
Finally we annihilate the pair of $\mu_1$ and $\bar{\mu}_1$ in the end \cite{Jiang:2014ksa}.
\item
The second description is that we take a Hopf link of $\mu_2$-${\mu_3}$ linking spinning around the loop of $\mu_1$ \cite{Jian:2014vfa, Bi:2014vaa}.
We denote the Hopf link of $\mu_2$-${\mu_3}$ as Hopf$[\mu_3,\mu_2]$, denote its spinning as Spun[Hopf$[\mu_3,\mu_2]]$,
and denote its linking with the third $T^2$-worldsheet of $\mu_1$ as Link[Spun[Hopf$[\mu_3,\mu_2]],\mu_1]$.
Thus we can define ${\tL^{\text{Tri}}_{{\mu_3}, {\mu_2}, {\mu_1}}}  \equiv Z[S^4; \text{Link[Spun[Hopf}[\mu_3,\mu_2]],\mu_1]]$.
From the second description, we immediate see that
${\tL^{\text{Tri}}_{{\mu_3}, {\mu_2}, {\mu_1}}}$ as $Z[S^4; \text{Link[Spun[Hopf}[\mu_3,\mu_2]],\mu_1]]$
are symmetric under exchanging $\mu_2 \leftrightarrow \mu_3$, up to an overall conjugation due to the orientation of quasi-excitations.
\end{itemize}

We can view the spacetime $S^4$ as a $S^4=\R^4 + \{\infty \}$, the Cartesian coordinate $\R^4$ plus a point at the infinity $ \{\infty \}$.
Similar to the embedding of Ref.~\cite{Jian:2014vfa}, we embed the $T^2$-worldsheets ${\mu_1}, {\mu_2}, {\mu_3}$ into the 
$(X_1,X_2,X_3,X_4) \in \R^4$ as follows:
\bea
\left\{
    \begin{array}{l}
   X_1(u, \vec{x})=[r_1(u)+(r_2(u)+r_3(u) \cos x) \cos y] \cos z,\\
   X_2(u, \vec{x})=[r_1(u)+(r_2(u)+r_3(u) \cos x) \cos y] \sin z,\\
   X_3(u, \vec{x})=(r_2(u)+r_3(u) \cos x) \sin y ,\\
   X_4(u, \vec{x})=r_3(u) \sin x,  \end{array}
    \right.\;\;\;\;
\eea
here $\vec{x}\equiv(x,y,z)$.
We choose the $T^2$-worldsheets as follows: \\
\noindent
The $T^2$-worldsheet ${\mu_1}$ is parametrized by some fixed $u_1$ and free coordinates of $(z,x)$ while $y=0$ is fixed.\\
\noindent
The $T^2$-worldsheet ${\mu_2}$ is parametrized by some fixed $u_2$ and free coordinates of $(x,y)$ while $z=0$ is fixed.\\
\noindent
The $T^2$-worldsheet ${\mu_3}$ is parametrized by some fixed $u_3$ and free coordinates of $(y,z)$ while $x=0$ is fixed.\\
We can set the parameters $u_1 > u_2 > u_3$. 
Meanwhile, a $T^3$-surface can be defined as
$\cM^3(u, \vec{x}) \equiv (X_1(u, \vec{x}),X_2(u, \vec{x}),X_3(u, \vec{x}),X_4(u, \vec{x}))$ with a fixed $u$ and free parameters $\vec{x}$.
The $T^3$-surface $\cM^3(u, \vec{x}) \equiv (X_1(u, \vec{x}),X_2(u, \vec{x}),X_3(u, \vec{x}),X_4(u, \vec{x}))$ encloses a 4-dimensional volume.
We define the enclosed 4-dimensional volume as the $\cM^3(u, \vec{x}) \times I^1(s)$ where $I^1(s)$ is the 1-dimensional radius interval along $r_3$,
such that
$I^1(s)=\{ s| s=[0, r_3(u)]\}$, namely $0 \leq s \leq r_3(u)$. Here we can define
$r_3(0)=0$. 
The topology of the enclosed 4-dimensional volume of $\cM^3(u, \vec{x}) \times I^1(s)$
 is of course the $T^3 \times I^1=T^2 \times (S^1 \times I^1) =T^2 \times D^2$.
For a $\cM^3(u_{\text{large}}, \vec{x})\times I^1(s)$ prescribed by a fixed larger $u_{\text{large}}$ and free parameters $\vec{x}$, the $\cM^3(u_{\text{large}}, \vec{x})\times I^1(s)$ must 
enclose the 4-volume spanned by the past history of $\cM^3(u_{\text{small}}, \vec{x})\times I^1(s)$, for any $u_{\text{large}}>u_{\text{small}}$.
Here we set $u_1> u_2> u_3$. And
we also set $r_1(u) > r_2(u) > r_3(u)$ for any given $u$.

One can check that the three $T^2$-worldsheet ${\mu_1},{\mu_2}$ and ${\mu_3}$ indeed have the nontrivial \emph{triple-linking number} \cite{carter2004surfaces}. 
We can design the triple-linking number to be: 
\bea
\text{Tlk$(\mu_2,\mu_1,\mu_3)$=Tlk$(\mu_3,\mu_1,\mu_2)=0$,
Tlk$(\mu_1,\mu_2,\mu_3)=+1$,} \nn\\
\text{Tlk$(\mu_3,\mu_2,\mu_1)=-1$, 
Tlk$(\mu_2,\mu_3,\mu_1)=+1$, Tlk$(\mu_1,\mu_3,\mu_2)=-1$.} 
\eea 

Below we will frequently use the surgery trick by cutting out a tubular neighborhood $D^2 \times T^2$ of the $T^2$-worldsheet
and re-gluing this $D^2 \times T^2$ back to its complement $S^4 \- D^2 \times T^2$ via the
modular $\cS^{xyz}$-transformation. The $\cS^{xyz}$-transformation sends
\bea
\bpm x_{\text{out}} \\
y_{\text{out}}  \\
z_{\text{out}} \epm=\bpm 0& 0&1 \\1& 0&0 \\0& 1&0\epm 
\bpm
x_{\text{in}} \\
y_{\text{in}}  \\
z_{\text{in}}\epm.
\eea 
Thus, the $\cS^{xyz}$-identification is 
$(x_{\text{out}},
y_{\text{out}},
z_{\text{out}}) \leftrightarrow 
(z_{\text{in}},
x_{\text{in}},
y_{\text{in}})$.
The $(\cS^{xyz})^{-1}$-identification is\\ 
$(x_{\text{out}},
y_{\text{out}},
z_{\text{out}})$ $\leftrightarrow$ 
$(y_{\text{in}},
z_{\text{in}},
x_{\text{in}})$.
The surgery on the initial $S^4$ outcomes a new manifold,
\be
(D^2 \times T^2) \cup_{T^3; \cS^{xyz}} (S^4 \- D^2 \times T^2) = S^3 \times S^1 \# S^2 \times S^2.
\ee

In terms of the spacetime path integral picture, use Eqs.(\ref{eq:surger1}) and (\ref{eq:surger2}), we derive:
\bea
&&Z \bpm \includegraphics[scale=0.45]{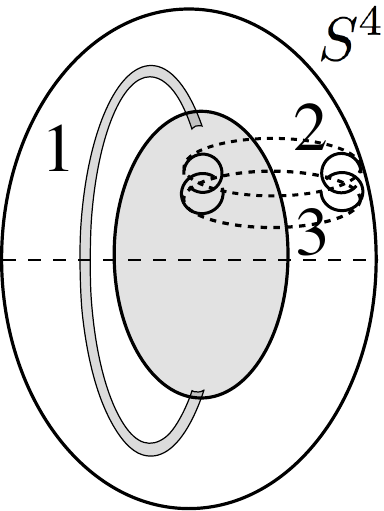} \epm \equiv {\tL^{\text{Tri}}_{{\mu_3}, {\mu_2}, {\mu_1}}}=Z[S^4; \text{Link[Spun[Hopf}[\mu_3,\mu_2]],\mu_1]] \nonumber\\
&&=\sum_{\mu_3'} {\cS^{xyz}_{\mu_3', \mu_3}} \; {Z(S^3 \times S^1 \# S^2 \times S^2; \mu_1, \mu_2 \parallel \mu_3')}  \label{eq:Sxyzsurger1} \\
&&=\sum_{\mu_3', {\Gamma_2}} {\cS^{xyz}_{\mu_3', \mu_3}} (\cF^{T^2})_{\mu_2 \mu_3'}^{\Gamma_2}  \; {Z(S^3 \times S^1 \# S^2 \times S^2; \mu_1, \Gamma_2)} \label{eq:Sxyzsurger2}\\
&&=\sum_{\mu_3', {\Gamma_2}, {\Gamma_2'}} {\cS^{xyz}_{\mu_3', \mu_3}} (\cF^{T^2})_{\mu_2 \mu_3'}^{\Gamma_2} {(\cS^{xyz})^{-1}_{\Gamma_2', \Gamma_2}} \; {Z(S^4; \mu_1, \Gamma_2')} \label{eq:Sxyzsurger3}\\
&&=\sum_{\mu_3', {\Gamma_2}, {\Gamma_2'}, {\Gamma_2''}} {\cS^{xyz}_{\mu_3', \mu_3}} (\cF^{T^2})_{\mu_2 \mu_3'}^{\Gamma_2} {(\cS^{xyz})^{-1}_{\Gamma_2', \Gamma_2}} 
{(\cS^{xyz})^{-1}_{\Gamma_2'', \Gamma_2'}} \; {Z(S^3 \times S^1 \# S^2 \times S^2;  \mu_1, \Gamma_2'')} \label{eq:Sxyzsurger4}\\
&&=\sum_{\mu_3', {\Gamma_2}, {\Gamma_2'}, {\Gamma_2''}, {\eta_2}} {\cS^{xyz}_{\mu_3', \mu_3}} (\cF^{T^2})_{\mu_2 \mu_3'}^{\Gamma_2} {(\cS^{xyz})^{-1}_{\Gamma_2', \Gamma_2}} 
{(\cS^{xyz})^{-1}_{\Gamma_2'', \Gamma_2'}}  (\cF^{T^2})_{{\mu_1} {\Gamma_2''}}^{\eta_2}\; {Z(S^3 \times S^1 \# S^2 \times S^2;  \eta_2)} \label{eq:Sxyzsurger5}
\quad\\
&&=\sum_{\mu_3', {\Gamma_2}, {\Gamma_2'}, {\Gamma_2''}, {\eta_2}, {\eta_2'}} {\cS^{xyz}_{\mu_3', \mu_3}} (\cF^{T^2})_{\mu_2 \mu_3'}^{\Gamma_2} {(\cS^{xyz})^{-1}_{\Gamma_2', \Gamma_2}} 
{(\cS^{xyz})^{-1}_{\Gamma_2'', \Gamma_2'}}  (\cF^{T^2})_{{\mu_1} {\Gamma_2''}}^{\eta_2}  {\cS^{xyz}_{\eta_2', \eta_2}}  \; {\tL^{\text{Tri}}_{0, 0, {\eta_2'}}} \label{eq:Sxyzsurger6}. 
\eea
As usual, the repeated indices are summed over. 
With the trick of $\cS^{xyz}$-transformation in mind, here is the step-by-step sequence of surgeries we perform.\\  


\noindent {\bf Step 1}:
We cut out the tubular neighborhood $D^2 \times T^2$ of the $T^2$-worldsheet of $\mu_3$ and re-glue this $D^2 \times T^2$ back to its complement $S^4 \- D^2 \times T^2$ via the
modular $(\cS^{xyz})^{-1}$-transformation. The $D^2 \times T^2$ neighborhood of $\mu_3$-worldsheet can be viewed as the 4-volume 
$\cM^3(u_{3}, \vec{x})\times I^1(s)$, which encloses neither $\mu_1$-worldsheet nor $\mu_2$-worldsheet. 
The $(\cS^{xyz})^{-1}$-transformation sends $(y_{\text{in}},
z_{\text{in}})$ of $\mu_3$ to
$(x_{\text{out}},
y_{\text{out}})$ of $\mu_2$. The gluing however introduces the summing-over new coordinate $\mu_3'$, 
based on Eq.(\ref{eq:surger1}). Thus Step 1 obtains Eq.(\ref{eq:Sxyzsurger1}).\\

In Step 1, as Eq.(\ref{eq:Sxyzsurger1}) and thereafter, we write down ${\cS^{xyz}_{\mu_3', \mu_3}}$ matrix.
Based on Eq.(\ref{eq:projectK}), 
we stress that the ${\cS^{xyz}_{\mu_3', \mu_3}}$ is projected to the $|0_{D^2 \times T^2} \rangle$-states with operator-insertions for both bra and ket states.\\ 
\bea
\label{eq:projectS}
&&{\cS^{xyz}_{\mu_3', \mu_3}} \equiv \< {\mu_3'}_{D^2 \times T^2}|  \hat{\cS}^{xyz} | {\mu_3}_{D^2 \times T^2} \rangle_{ {D^2 \times T^2} \cup_{T^3; \hat{S}^{xyz}} {D^2 \times T^2}} \nonumber \\
&&=\< 0_{D^2_{xw} \times T^2_{yz}} | V^{T^2_{yz} \dagger}_{\mu_3'}  \hat{\cS}^{xyz}  V^{T^2_{yz}}_{\mu_3}  | 0_{D^2_{xw} \times T^2_{yz}} \rangle_{S^3 \times S^1} 
\eea
Here we use the surgery fact 
\be
{ {D^2 \times T^2} \cup_{T^3; \hat{S}^{xyz}} {D^2 \times T^2}}={S^3 \times S^1}.
\ee
So our ${\cS^{xyz}_{\mu_3', \mu_3}}$ is defined as a quantum amplitude in ${S^3 \times S^1}$. Two $T^2$-worldsheets
${\mu_3'}$ and ${\mu_3}$ now become a pair of Hopf link resides in $S^3$ part of ${S^3 \times S^1}$, while
share the same $S^1$ circle in the $S^1$ part of ${S^3 \times S^1}$.
We can view the shared $S^1$ circle as the spinning circle of the spun surgery construction 
on the Hopf link in $D^3$, the spun-topology would be
$D^3 \times S^1$, then we glue this $D^3 \times S^1$ contains $\text{Spun[Hopf}[{\mu_3'},{\mu_3}]]$ to another $D^3 \times S^1$, so 
we have $D^3 \times S^1 \cup_{S^2 \times S^1} D^3 \times S^1= S^3 \times S^1$ as an overall new spacetime topology.
Hence we also denote 
\be
{\cS^{xyz}_{\mu_3', \mu_3}}=Z[{S^3 \times S^1}; \text{Spun[Hopf}[{\mu_3'},{\mu_3}]]]. 
\ee

\noindent {\bf Step 2:} 
The earlier surgery now makes the inner $\mu_3'$-worldsheet \emph{parallels} to the outer $\mu_2$-worldsheet,
since they share the same coordinates $(x_{\text{out}}, y_{\text{out}})=(y_{\text{in}},z_{\text{in}})$. We denote their parallel topology as $\mu_2 \parallel \mu_3'$.
So we can fuse 
the $\mu_2$-worldsheet and $\mu_3'$-worldsheet via the fusion algebra, namely
$V^{T^2_{x_{\text{out}},
y_{\text{out}}}}_{\mu_2} V_{\mu_3'}^{T^2_{x_{\text{out}},
y_{\text{out}}}}=  (\cF^{T^2})_{\mu_2 \mu_3'}^{\Gamma_2} V_{\Gamma_2}^{T^2_{x_{\text{out}},
y_{\text{out}}}}$. Thus Step 2 obtains Eq.(\ref{eq:Sxyzsurger2}). \\

\noindent {\bf Step 3:} 
We cut out the tubular neighborhood $D^2 \times T^2$ of the $T^2$-worldsheet of ${\Gamma_2}$ and re-glue this $D^2 \times T^2$ back to its complement $S^4 \- D^2 \times T^2$ via the
modular $\cS^{xyz}$-transformation. The $D^2 \times T^2$ neighborhood of ${\Gamma_2}$-worldsheet can be viewed as the 4-volume 
$\cM^3(u_{2}, \vec{x})\times I^1(s)$ in the new manifold $S^3 \times S^1 \# S^2 \times S^2$, which encloses no worldsheet inside. 
After the surgery, 
the $\cS^{xyz}$-transformation sends the redefined $(x_{\text{in}},
y_{\text{in}})$ of $\Gamma_2$ back to
$(y_{\text{out}},
z_{\text{out}})$ of $\Gamma_2'$. The gluing however introduces the summing-over new coordinate $\Gamma_2'$, 
based on Eq.(\ref{eq:surger1}). 
We also transform $S^3 \times S^1 \# S^2 \times S^2$ back to $S^4$ again. 
Thus Step 3 obtains Eq.(\ref{eq:Sxyzsurger3}).\\

\noindent {\bf Step 4:} 
We cut out the tubular neighborhood $D^2 \times T^2$ of the $T^2$-worldsheet of ${\Gamma_2'}$ and re-glue this $D^2 \times T^2$ back to its complement $S^4 \- D^2 \times T^2$ via the
modular $\cS^{xyz}$-transformation. The $D^2 \times T^2$ neighborhood of ${\Gamma_2}'$-worldsheet viewed as the 4-volume 
in the manifold $S^4$ encloses no worldsheet inside. 
After the surgery, 
the $\cS^{xyz}$-transformation sends the $(x_{\text{in}},
y_{\text{in}})$ of $\Gamma_2'$ to
$(z_{\text{out}},
x_{\text{out}})$ of $\mu_1$. The gluing however introduces the summing-over new coordinate $\Gamma_2''$, 
based on Eq.(\ref{eq:surger1}). 
We also transform $S^4$ to  $S^3 \times S^1 \# S^2 \times S^2$ again. 
Thus Step 4 obtains Eq.(\ref{eq:Sxyzsurger4}).\\

\noindent {\bf Step 5:} 
The earlier surgery now makes the inner $\Gamma_2''$-worldsheet \emph{parallels} to the outer $\mu_1$-worldsheet,
since they share the same coordinates $(z_{\text{out}},
x_{\text{out}})=(x_{\text{in}},
y_{\text{in}})$. We denote their parallel topology as $\mu_1 \parallel \Gamma_2''$.
We now fuse 
the $\mu_1$-worldsheet and $\Gamma_2''$-worldsheet via the fusion algebra, namely
$V_{\mu_1}^{T^2_{z_{\text{out}},
x_{\text{out}}}} V_{\Gamma_2''}^{T^2_{z_{\text{out}},
x_{\text{out}}}}=  (\cF^{T^2})_{{\mu_1} {\Gamma_2''}}^{\eta_2} V_{\eta_2}^{T^2_{z_{\text{out}},
x_{\text{out}}}}$. Thus Step 5 obtains Eq.(\ref{eq:Sxyzsurger5}). \\ 

\noindent {\bf Step 6:} 
We should do the inverse transformation to get back to the $S^4$ manifold. 
Thus we cut out the tubular neighborhood $D^2 \times T^2$ of the $T^2$-worldsheet of ${\eta_2}$ and re-glue this $D^2 \times T^2$ back to its complement via the
modular $(\cS^{xyz})^{-1}$-transformation. We relate the original path integral to the final one ${Z(S^4;  \eta_2')}={\tL^{\text{Tri}}_{0, 0, {\eta_2'}}}$.
  Thus Step 6 obtains Eq.(\ref{eq:Sxyzsurger6}).

Similarly, in the fourth path integral of Eq.(\ref{eq:Spin[HopfLink]S4}), we derive
\bea \label{eq:SxyzsurgerDown}
&&Z \bpm \includegraphics[scale=0.45]{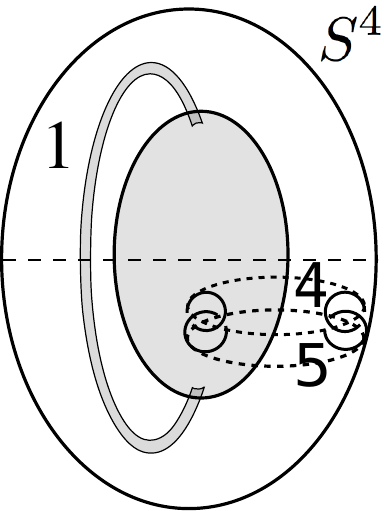} 
\epm \equiv {\tL^{\text{Tri}}_{{\mu_5}, {\mu_4}, {\mu_1}}}=Z[S^4; \text{Link[Spun[Hopf}[\mu_5,\mu_4]],\mu_1]] \nonumber\\
&&=\sum_{\mu_5', {\Gamma_4}, {\Gamma_4'}, {\Gamma_4''}, {\eta_4}, {\eta_4'}} {\cS^{xyz}_{\mu_5', \mu_5}} (\cF^{T^2})_{\mu_4 \mu_5'}^{\Gamma_4} {(\cS^{xyz})^{-1}_{\Gamma_4', \Gamma_4}} 
{(\cS^{xyz})^{-1}_{\Gamma_4'', \Gamma_4'}}  (\cF^{T^2})_{{\mu_1} {\Gamma_4''}}^{\eta_4}  {\cS^{xyz}_{\eta_4', \eta_4}}  \; {\tL^{\text{Tri}}_{0, 0, {\eta_4'}}}. 
\eea

In the second path integral of Eq.(\ref{eq:Spin[HopfLink]S4}), 
we have the Hopf link of $\text{Hopf}[\mu_3,\mu_2]$ and the Hopf link of $\text{Hopf}[\mu_5,\mu_4]$. 
In the spacetime picture, all $\mu_2, \mu_3, \mu_4, \mu_5$ are $T^2$-worldsheets under the spun surgery construction. 
We can locate the the spun object named $\text{Spun[Hopf}[\mu_3,\mu_2],\text{Hopf}[\mu_5,\mu_4]]$ inside a $D^3 \times S^1$, while this
$D^3 \times S^1$ is glued with a $S^2 \times D^2$ to a $S^4$. Here the $S^2 \times D^2$ contains a $T^2$-worldsheet $\mu_1$.
We can view the $T^2$-worldsheet $\mu_1$ contains a $S^2$-sphere of the $S^2 \times D^2$ but attached an extra handle.
We derive: 
\bea
&&Z \bpm \includegraphics[scale=0.45]{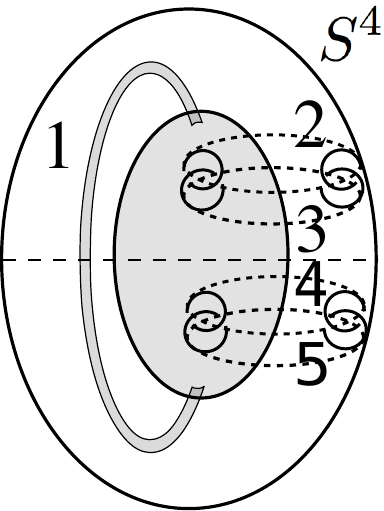} 
\epm \equiv Z[S^4; \text{Link[Spun[Hopf}[\mu_3,\mu_2],\text{Hopf}[\mu_5,\mu_4]]],\mu_1]] \nonumber\\
&&=
\sum_{\mu_3', {\Gamma_2}, {\Gamma_2'}}
\sum_{\mu_5', {\Gamma_4}, {\Gamma_4'}}
{\cS^{xyz}_{\mu_3', \mu_3}} (\cF^{T^2})_{\mu_2 \mu_3'}^{\Gamma_2} {(\cS^{xyz})^{-1}_{\Gamma_2', \Gamma_2}} 
{\cS^{xyz}_{\mu_5', \mu_5}} (\cF^{T^2})_{\mu_4 \mu_5'}^{\Gamma_4} {(\cS^{xyz})^{-1}_{\Gamma_4', \Gamma_4}}
Z[S^4; \text{Spun}[\Gamma_2',\Gamma_4'],\mu_1] \label{eq:DoubleHopf1}  \quad\\
&&=
\sum_{\mu_3', {\Gamma_2}, {\Gamma_2'}}
\sum_{\mu_5', {\Gamma_4}, {\Gamma_4'}} \sum_{\Gamma}
{\cS^{xyz}_{\mu_3', \mu_3}} (\cF^{T^2})_{\mu_2 \mu_3'}^{\Gamma_2} {(\cS^{xyz})^{-1}_{\Gamma_2', \Gamma_2}} 
{\cS^{xyz}_{\mu_5', \mu_5}} (\cF^{T^2})_{\mu_4 \mu_5'}^{\Gamma_4} {(\cS^{xyz})^{-1}_{\Gamma_4', \Gamma_4}}
(\cF^{T^2})_{\Gamma_2',\Gamma_4'}^{\Gamma} Z[S^4; \Gamma,\mu_1] \label{eq:DoubleHopf2} \;\;\;  \quad\quad\\
&&=\sum_{\overset{\mu_3', {\Gamma_2}, {\Gamma_2'}}{\mu_5', {\Gamma_4}, {\Gamma_4'}}}
\sum_{\Gamma, \Gamma', \Gamma_1, \Gamma_1'}
{\cS^{xyz}_{\mu_3', \mu_3}} (\cF^{T^2})_{\mu_2 \mu_3'}^{\Gamma_2} {(\cS^{xyz})^{-1}_{\Gamma_2', \Gamma_2}} 
{\cS^{xyz}_{\mu_5', \mu_5}} (\cF^{T^2})_{\mu_4 \mu_5'}^{\Gamma_4} {(\cS^{xyz})^{-1}_{\Gamma_4', \Gamma_4}} \nn\\
&&\quad\quad\quad\quad\quad\quad\quad\quad\quad\quad\quad\quad\quad\quad\quad \cdot (\cF^{T^2})_{\Gamma_2',\Gamma_4'}^{\Gamma}
{(\cS^{xyz})^{-1}_{\Gamma', \Gamma}}  (\cF^{T^2})_{{\mu_1} {\Gamma'}}^{\Gamma_1}  {\cS^{xyz}_{\Gamma_1', \Gamma_1}}  \; {\tL^{\text{Tri}}_{0, 0, {\Gamma_1'}}}. \label{eq:DoubleHopf3}\;\;\;\;\;\;\;\;\;
\eea
Here we do the Step 1, Step 2 and Step 3 surgeries on $\text{Spun[Hopf}[\mu_3,\mu_2]]$ first, then do the same 3-step surgeries on $\text{Spun[Hopf}[\mu_5,\mu_4]]$ later,
then we obtain Eq.(\ref{eq:DoubleHopf1}).
While in Eq.(\ref{eq:DoubleHopf1}), the new $T^2$-worldsheets ${\Gamma_2'}$ and ${\Gamma_4'}$ have no triple-linking with the worldsheet $\mu_1$. 
Here ${\Gamma_2'}$ and ${\Gamma_4'}$ are arranged in the $D^3 \times S^1$ part of the $S^4$ manifold, while $\mu_1$ is in the $S^2 \times D^2$ part of the $S^4$ manifold.
Indeed, ${\Gamma_2'}$ and ${\Gamma_4'}$ can be fused together in parallel to a new $T^2$-worldsheet ${\Gamma}$ via the fusion algebra $(\cF^{T^2})_{\Gamma_2',\Gamma_4'}^\Gamma$, so we obtain Eq.(\ref{eq:DoubleHopf2}).
Then we apply the Step 4, Step 5 and Step 6 surgeries on the $T^2$-worldsheets $\Gamma$ and $\mu_1$ of $Z[S^4; \text{Spun}[\Gamma],\mu_1]= Z[S^4; \Gamma,\mu_1]$ in Eq.(\ref{eq:DoubleHopf2}),
we obtain the final form Eq.(\ref{eq:DoubleHopf3}).

Use Eqs.(\ref{eq:S4T2}),(\ref{eq:Sxyzsurger6}),(\ref{eq:SxyzsurgerDown}) and (\ref{eq:DoubleHopf3}), and plug them into the path integral surgery relations, 
after massaging the relations, we derive a new quantum surgery formula (namely Eq.(\ref{eq:S2S1S1inS4}) in the earlier text):
\bea
\label{}
&&\ \ \ \
  Z \bpm \includegraphics[scale=0.45]{3+1D_Triple_link_T2_1.pdf} \epm 
 Z \bpm \includegraphics[scale=0.45]{3+1D_Triple_link_double_SpinHopfLink_S2_double_12345.pdf} \epm 
=
 Z \bpm \includegraphics[scale=0.45]{3+1D_Triple_link_double_SpinHopfLink_S2_Up_123.pdf} \epm 
 Z \bpm \includegraphics[scale=0.45]{3+1D_Triple_link_double_SpinHopfLink_S2_Down_145.pdf} \epm \nonumber \\
&& \Rightarrow
 {{\tL^{\text{Tri}}_{0, 0, {\mu_1}}} \cdot
 \sum_{\Gamma, \Gamma', \Gamma_1, \Gamma_1'}
 (\cF^{T^2})_{\Gamma_2',\Gamma_4'}^{\Gamma}
{(\cS^{xyz})^{-1}_{\Gamma', \Gamma}}  (\cF^{T^2})_{{\mu_1} {\Gamma'}}^{\Gamma_1}  {\cS^{xyz}_{\Gamma_1', \Gamma_1}}  \; {\tL^{\text{Tri}}_{0, 0, {\Gamma_1'}}} } \nonumber
\\
&&
{=
{\sum_{{\Gamma_2''}, {\eta_2}, {\eta_2'}} 
{(\cS^{xyz})^{-1}_{\Gamma_2'', \Gamma_2'}}  (\cF^{T^2})_{{\mu_1} {\Gamma_2''}}^{\eta_2}  {\cS^{xyz}_{\eta_2', \eta_2}}  \; {\tL^{\text{Tri}}_{0, 0, {\eta_2'}}} 
} 
\cdot
{\sum_{{\Gamma_4''}, {\eta_4}, {\eta_4'}}  
{(\cS^{xyz})^{-1}_{\Gamma_4'', \Gamma_4'}}  (\cF^{T^2})_{{\mu_1} {\Gamma_4''}}^{\eta_4}  {\cS^{xyz}_{\eta_4', \eta_4}}  \; {\tL^{\text{Tri}}_{0, 0, {\eta_4'}}}
 }}, \nn
\eea
here only ${\mu_1},{\Gamma_2'},{\Gamma_4'}$ are the fixed indices, other indices are summed over.
%

\subsubsection{More discussions}

For all path integrals of $S^4$ in Eqs.(\ref{eq:S2S1S1inS4}), (\ref{eq:S1S2S2inS4}) and (\ref{eq:Spin[HopfLink]S4}), each $S^4$ is cut into two $D^4$ pieces, 
so $D^4 \cup_{S^3} D^4={S^4}$. We choose all the dashed cuts for 3+1D path integral
representing $B=S^3$, while
we can view the $S^3$ as a spatial slice,
with the following excitation configurations:

\begin{enumerate}[label=\textcolor{blue}{(\roman*).}, ref={(\roman*)},leftmargin=*]

\item 
A loop in Eq.(\ref{eq:S2S1S1inS4}) on the slice $B=S^3$.
Thus, we consider a natural 1-dimensional Hilbert space (GSD=1),
for a single shrinkable loop excitation configuration on the spatial $S^3$.

\item
A pair of particle-antiparticle in Eq.(\ref{eq:S1S2S2inS4}) on the slice $B=S^3$. Thus, we consider a natural 1-dimensional Hilbert space (GSD=1),
for this particle-antiparticle configuration on the spatial $S^3$.

\item A pair of loop-antiloop in Eq.(\ref{eq:Spin[HopfLink]S4}) on the slice $B=S^3$.
In this case, here we require a stronger criterion that all loop excitations are gapped without zero modes, 
then the GSD is 1 for all above spatial section $S^3$. 

\end{enumerate}

Thus all our surgeries satisfy the assumptions for Eq.(\ref{ZMN}).

\subsubsection{Formulas for 3+1D fusion statistics}

The above Verlinde-like formulas 
constrain the fusion data (e.g. $\cN$, $\cF^{S^1}$, $\cF^{S^2}$, $\cF^{T^2}$, etc.)  
and braiding data (e.g. $\cS$, $\cT$, ${\tL}^\text{($S^2$,$S^1$)}$, ${\tL^{\text{Tri}}}$, $\cS^{xyz}$, etc.). 
Moreover, we can derive constraints between the fusion data itself.
Since a $T^2$-worldsheet contains two non-contractible $S^1$-worldlines along its two homology group generators in $H_1(T^2,\Z)=\Z^2$,
the $T^2$-worldsheet operator $V^{T^2}_{\mu}$ contains the data of $S^1$-worldline operator $W^{S^1}_{\si}$. 
More explicitly, we can compute the state $W_{\sigma_1}^{S^1_y}  W_{\sigma_2}^{S^1_y} V_{\mu_2}^{T^2_{yz}} | 0_{D^2_{wx} \times T^2_{yz}} \rangle$
by fusing two $W^{S^1}_{\si}$ operators and one $V^{T^2}_{\mu}$ operator in different orders, then we obtain a consistency formula: 
\bea \label{eq:FS1FT2}
\sum_{\sigma_3} (\cF^{S^1})_{{\sigma_1}{\sigma_2}}^{\sigma_3}   (\cF^{T^2})_{{\sigma_3}{\mu_2}}^{\mu_3}
=\sum_{\mu_1} (\cF^{T^2})_{{\sigma_2}{\mu_2}}^{\mu_1} (\cF^{T^2})_{{\sigma_1}{\mu_1}}^{\mu_3}. \;\;\;\;\; 
\eea
We organize 
our quantum statistics data of fusion and braiding, and some explicit examples of 
topological orders and their topological invariances in terms of our data in the Supplemental Material.


Lastly we provide more explicit calculations of Eq.~(\ref{eq:FS1FT2}), the constraint between the fusion data itself.
First, we recall that
\bea
&& W_{\sigma_1}^{S^1_y}  W_{\sigma_2}^{S^1_y} = (\cF^{S^1})_{{\sigma_1}{\sigma_2}}^{\sigma_3} W_{\sigma_3}^{S^1_y}, \nn\\
&&  V_{\mu_1}^{T^2_{yz}} V_{\mu_2}^{T^2_{yz}}=  (\cF^{T^2})_{{\mu_1}{\mu_2}}^{\mu_3} V_{\mu_3}^{T^2_{yz}}, \nn\\
&& W_{\sigma_1}^{S^1_y} V_{\mu_2}^{T^2_{yz}}=(\cF^{T^2})_{{\sigma_1}{\mu_2}}^{\mu_3} V_{\mu_3}^{T^2_{yz}}.\nn
\eea
Of course, the fusion algebra is symmetric respect to exchanging the lower indices, $(\cF^{T^2})_{{\sigma_1}{\mu_2}}^{\mu_3}=(\cF^{T^2})_{{\mu_2}{\sigma_1}}^{\mu_3}$.
We can regard the fusion algebra $(\cF^{S^1})_{{\sigma_1}{\sigma_2}}^{\sigma_3}$ and $(\cF^{T^2})_{{\sigma_1}{\mu_2}}^{\mu_3}$ with worldlines
as a part of a larger algebra of the fusion algebra of worldsheets $(\cF^{T^2})_{{\mu_1}{\mu_2}}^{\mu_3}$.
We compute the state $W_{\sigma_1}^{S^1_y}  W_{\sigma_2}^{S^1_y} V_{\mu_2}^{T^2_{yz}} | 0_{D^2_{wx} \times T^2_{yz}} \rangle$
by fusing two $W^{S^1}_{\si}$ operators and one $V^{T^2}_{\mu}$ operator in different orders.

On one hand, we can fuse two worldlines first, then fuse with the worldsheet,
\bea
&& W_{\sigma_1}^{S^1_y}  W_{\sigma_2}^{S^1_y} V_{\mu_2}^{T^2_{yz}} | 0_{D^2_{wx} \times T^2_{yz}} \rangle \nonumber \\
&&=\sum_{{\sigma_3}} (\cF^{S^1})_{{\sigma_1}{\sigma_2}}^{\sigma_3} W_{\sigma_3}^{S^1_y} V_{\mu_2}^{T^2_{yz}} | 0_{D^2_{wx} \times T^2_{yz}} \rangle \nonumber\\
&&=\sum_{{\sigma_3},{\mu_3}} (\cF^{S^1})_{{\sigma_1}{\sigma_2}}^{\sigma_3}   (\cF^{T^2})_{{\sigma_3}{\mu_2}}^{\mu_3}  V_{\mu_3}^{T^2_{yz}} | 0_{D^2_{wx} \times T^2_{yz}} \rangle. \label{eq:WWV1}
\eea
On the other hand, we can fuse a worldline with the worldsheet first, then fuse with another worldline,
\bea
&& W_{\sigma_1}^{S^1_y}  W_{\sigma_2}^{S^1_y} V_{\mu_2}^{T^2_{yz}} | 0_{D^2_{wx} \times T^2_{yz}} \rangle \nonumber\\
&&=\sum_{{\mu_1}} W_{\sigma_1}^{S^1_y} (\cF^{T^2})_{{\sigma_2}{\mu_2}}^{\mu_1}  V_{\mu_1}^{T^2_{yz}}  | 0_{D^2_{wx} \times T^2_{yz}} \rangle \nonumber\\
&&=\sum_{{\mu_1},{\mu_3}} (\cF^{T^2})_{{\sigma_2}{\mu_2}}^{\mu_1} (\cF^{T^2})_{{\sigma_1}{\mu_1}}^{\mu_3}
V_{\mu_3}^{T^2_{yz}}  | 0_{D^2_{wx} \times T^2_{yz}} \rangle \label{eq:WWV2}
\eea
Therefore, by comparing Eqs.(\ref{eq:WWV1}) and (\ref{eq:WWV2}),
we derive a consistency condition for fusion algebra Eq.~(\ref{eq:FS1FT2}):
$
\sum_{\sigma_3} (\cF^{S^1})_{{\sigma_1}{\sigma_2}}^{\sigma_3}   (\cF^{T^2})_{{\sigma_3}{\mu_2}}^{\mu_3}
=\sum_{\mu_1} (\cF^{T^2})_{{\sigma_2}{\mu_2}}^{\mu_1} (\cF^{T^2})_{{\sigma_1}{\mu_1}}^{\mu_3}. \;\;\;\;\; 
$

\section{Summary of Quantum Statistics Data of Fusion and Braiding}

\begin{table}[!h]
\begin{tabular}{l}
\hline
\hline\\[-2mm]
Quantum statistics data of fusion and braiding\\[1mm]
\hline
\hline\\[-2mm]
Data for {\bf 2+1D} topological orders:\\[1mm]
\hline\\[-2mm]
$\bullet$ Fusion data:\\[1mm]
{$\cN^{\si_3}_{\si_1 \si_2}=\cF^{\si_3}_{\si_1 \si_2}$} (fusion tensor) from \eqn{eq:Nabc},\\[1mm] 
$\bullet$ Braiding data:\\[1mm]
{$\cS^{xy}$, $\cT^{xy}$} (modular SL$(2,Z)$ matrices from MCG$(T^2)$) from \eqn{S12} and \eqn{T12},\\[1mm]
$Z[T^3_{xyt};\sigma'_{1x},\sigma'_{2y},\sigma'_{3t}]$ (or $Z[S^3; \text{BR}[\sigma_1,\sigma_2,\sigma_3]$) from \eqn{SBR}, etc.\\[1mm]
\hline\\[-2mm]
Data for {\bf 3+1D} topological orders:\\[1mm]
\hline\\[-2mm]
$\bullet$ Fusion data:\\[1mm]
{$(\cF^{S^1})^{\si_3}_{\si_1 \si_2}$, $(\cF^{S^2})^{\mu_3}_{\mu_1 \mu_2}$, $(\cF^{T^2})^{\mu_3}_{\mu_1 \mu_2}$. } (fusion tensor) from
\eqn{F3+1DS1}, \eqn{F3+1DS2}, and \eqn{F3+1DT2}. \\[1mm]
$\bullet$ Braiding data:\\[1mm]
{$\cS^{xyz}$, $\cT^{xy}$} (modular SL$(3,\Z)$ matrices from MCG$(T^3)$ from \eqn{eq:Sxyz} and \eqn{eq:Txy}, \\[1mm]
including {$\cS^{xy}$})\\[1mm]
$\tL^{\text{Tri}}_{{0,0,\mu}}$ (from ${ \tL^{\text{Tri}}_{\mu_3,\mu_2,\mu_1}}$ of \eqn{T2T2T2glue}),  {$\tL^\text{Lk($S^2$,$S^1$)}_{\mu \si}$} from \eqn{S2S1glue}, \\[1mm]
$Z[T^4 \# S^2  \times S^2; \mu_4',\mu_3',\mu_2',\mu_1' ]$ from \eqn{eq:ZT4S2}\\[1mm] 
(from $Z[S^4; \text{Link[Spun[BR}[\mu_4 ,\mu_3, \mu_2]],\mu_1]]$ of \eqn{ZSpinBRS4}), etc. \\[1mm]
\hline
\end{tabular}
\caption{Some data for 2+1D and 3+1D topological orders encodes their quantum statistics properties, such as fusion and braiding statistics of their quasi-excitations (anyonic particles and anyonic strings).
However, the data is not complete because we do not account the degrees of freedom of their boundary modes, such as the chiral central charge $c_-=c_L-c_R$ for 2+1D  topological orders.}
\label{table:data}
\end{table}

We organize the quantum statistics data of fusion and braiding introduced in the earlier text into Table \ref{table:data}.
We propose using the set of data in Table \ref{table:data} to label or characterize topological orders, although such labels may only partially characterize topological orders. 
We also remark that
Table \ref{table:data} may not contain all sufficient data to characterize and classify all topological orders. 
What can be the missing data in Table \ref{table:data}?
Clearly, there is the chiral central charge $c_-=c_L-c_R$, the difference between the left and right central charges, missing for 2+1D  topological orders.
The $c_-$ is essential for describing 2+1D topological orders with 1+1D boundary gapless chiral edge modes.
The gapless chiral edge modes cannot be fully gapped out by adding scattering terms between different modes, because they are protected by the net chirality.
So our 2+1D data only describes \emph{2+1D non-chiral topological orders}.
Similarly, our 2+1D/3+1D data may not be able to fully classify {2+1D/3+1D topological orders} whose boundary modes are \emph{protected to be gapless}. 
We may need additional data to encode boundary degrees of freedom for their boundary modes.

In some case, some of our data may overlap with the information given by other data. For example, the 2+1D topological order data
($\cS_{xy}, \cT_{xy}$, $\cN^{\si_3}_{\si_1 \si_2}$) may contain the information of $Z[T^3_{xyt};\sigma'_{1x},\sigma'_{2y},\sigma'_{3t}]$ (or $Z[S^3; \text{BR}[\sigma_1,\sigma_2,\sigma_3]$) already,
since we know that we the former set of data may fully classify 2+1D bosonic topological orders.

Although it is possible that there are extra required data beyond what we list in Table \ref{table:data}, we find that Table \ref{table:data} is sufficient enough for
a large class of topological orders, at least for those described by Dijkgraaf-Witten twisted gauge theory \cite{Dijkgraaf:1989pz} and those gauge theories with finite Abelian gauge groups.
In the next Section, 
we will give some explicit examples of 2+1D and 3+1D topological orders described by 
Dijkgraaf-Witten topological gauge theory and group cohomology \cite{Dijkgraaf:1989pz}, which can be completely characterized and classified by the data given in Table \ref{table:data}.

\section{Examples of Topological Orders, TQFTs and Topological Invariances 
\label{sec:example}}

In Table \ref{table:TQFT}, 
we give some explicit examples of 2+1D and 3+1D topological orders from Dijkgraaf-Witten twisted gauge theory.
We like to emphasize that our quantum-surgery Verlinde-like formulas apply to generic  2+1D and 3+1D topological orders beyond the gauge theory or field theory description.
So our formulas apply to quantum phases of matter or theories beyond the Dijkgraaf-Witten twisted gauge theory description. 
We list down these examples only because these are famous examples with a more familiar gauge theory understanding.
In terms of topological order language, Dijkgraaf-Witten theory describes the low energy physics of certain bosonic topological orders which 
can be regularized on a lattice Hamiltonian \cite{{Wang:2014oya},{Jiang:2014ksa},{Wan:2014woa}} with local bosonic degrees of freedom (without local fermions, but there can be emergent fermions and anyons).

We also clarify that what we mean by the correspondence between the items in the same row in Table \ref{table:TQFT}:
%
\begin{itemize}
\item
\cblue{(i)} 
Quantum statistic braiding data, 
\item 
\cblue{(ii)} Group cohomology cocycles 
\item 
\cblue{(iii)} 
Topological quantum field theory (TQFT). 
\end{itemize}
What we mean is that we can distinguish
the topological orders of given cocycles of (ii) with the low energy TQFT of (iii) by measuring their quantum statistic Berry phase 
under the prescribed braiding process in the path integral of (i). 
\cblue{The Euclidean
path integral of (i) is defined through the action $\mathbf{S}$ of (iii)
via
\be
Z=\int [DB_I][DA_I]\exp[ - \mathbf{S}].
\ee
}
For example, the mutual braiding (Hopf linking) measures the $\cS$ matrix
distinguishing different types of  $\int \frac{\ii N_I}{2\pi}{B^I  \wedge d A^I} + { \frac{{\ii} p_{IJ}}{2 \pi}} A^I \wedge dA^J$ with different $p_{IJ}$ couplings;
while the Borromean ring braiding can distinguish different types of $\int \frac{\ii N_I}{2\pi}{B^I  \wedge d A^I}+{{\ii} c_{123}} A^1 \wedge A^2 \wedge  A^3$
with different $c_{123}$ couplings. However, the table does not mean that we cannot use braiding data in one row to measures the TQFT in another row.
For example, 
$\cS$ matrix can also distinguish the $\int \frac{\ii N_I}{2\pi}{B^I  \wedge d A^I}+{{\ii} c_{123}} A^1 \wedge A^2 \wedge  A^3$-type theory.
However, $Z[S^3; \text{BR}[\sigma_1,\sigma_2,\sigma_3]=Z[T^3_{xyt};\sigma'_{1x},\sigma'_{2y},\sigma'_{3t}]=1$ is trivial for 
$\int \frac{\ii N_I}{2\pi}{B^I  \wedge d A^I} + { \frac{{\ii} p_{IJ}}{2 \pi}} A^I \wedge dA^J$ with any $p_{IJ}$. Thus Borromean ring braiding cannot
measure nor distinguish the nontrivial-ness of $p_{IJ}$-type theories.

Recently, after the appearance of our previous work \cite{JWangthesis, Wang2016qhf1602.05951}, 
further progress has been made on systematically and rigorously deriving the topological invariants of TQFTs, such as: 
\begin{enumerate}[label=\textcolor{blue}{\arabic*.}, ref={\arabic*},leftmargin=*]
\item 
The TQFT link invariants \cite{Putrov2016qdo1612.09298}, 
\item
The GSD data and the partition function \cite{Wang2018edf1801.05416} without extended operator insertions, with or without topological boundary, 
\end{enumerate}
directly from
the continuum bosonic TQFT formulation.

\newgeometry{left=1.2cm, right=1.4cm, top=0.4cm, bottom=2.cm}
\begin{center}
\begin{table}[!h]
\noindent
\fontsize{9}{9}\selectfont
\hspace*{9mm}
{
\makebox[\textwidth][r]{
\begin{tabular}{ccc} 
\hline
$\begin{matrix}
\text{(i). Path-integral linking 
invariants;}\\
\text{Quantum statistic braiding data} 
\end{matrix}$ & 
$\begin{matrix}\text{(ii). Group-cohomology cocycles}\\ 
\text{distinguished by the braiding in (i)} 
\end{matrix}$ & 
$\begin{matrix}\text{(iii). TQFT actions \cblue{$\mathbf{S}$} characterized}\\ \text{by the spacetime-braiding in (i)} 
\end{matrix}$ \\
\hline\\[-2mm]
\multicolumn{3}{c}{2+1D} \\
\cline{1-3}\\[-2mm]
$\begin{matrix}
Z \bpm \includegraphics[scale=0.3]{S3ll12_uncut_l.pdf} \includegraphics[scale=0.24]{S3_top.pdf}\epm  \\   
=Z[S^3; \text{Hopf}[\sigma_1,\sigma_2]] 
=\cS^{}_{\bar{\sigma}_1\sigma_2}
 \end{matrix}$
& $\exp \Big( \frac{2 \pi \ii {p_{IJ} }  }{N_{I} N_{J}} \; a_{I}(b_{J} +c_{J} -[b_{J} +c_{J}]) \Big) $  &  
$\begin{matrix}
\ii \int 
\frac{ N_I}{2\pi}{B^I  \wedge \dd A^I} + { \frac{ p_{IJ}}{2 \pi}} A^I \wedge \dd A^J\\[2mm]
A^I \to A^I+ \dd g^I, \\[2mm]
N_I B^I  \to N_I B^I + \dd \eta^I.
 \end{matrix}
$  \\
\hline\\[-2mm]
$\begin{matrix}
Z \bpm \includegraphics[scale=0.35]{2+1D_Borromean_mid_123_l.pdf} \epm\\
=Z[S^3; \text{BR}[\sigma_1,\sigma_2,\sigma_3];\\
\text{Also } Z[T^3_{xyt};\sigma'_{1x},\sigma'_{2y},\sigma'_{3t}]
 \end{matrix}$      &   $ \exp \Big( \frac{2 \pi \ii p_{123}  }{N_{123}} \;  a_{1}b_{2}c_{3} \Big)$   & 
$\begin{matrix}
\ii \int 
\frac{ N_I}{2\pi}{B^I  \wedge \dd A^I}+{{} c_{123}} A^1 \wedge A^2 \wedge  A^3\\[2mm]
A^I \to A^I+ \dd g^I, \\[2mm]
N_I B^I  \to N_I B^I + \dd \eta^I+ 2\pi {\tilde{c}}_{IJK}  A^J  g^K \\[2mm] 
- \pi {\tilde{c}}_{IJK}  g^J \dd g^K.
\end{matrix}$      \\
\hline\\[-2mm]
\multicolumn{3}{c}{3+1D} \\
\cline{1-3}\\[-2mm]
$\begin{matrix}
Z \bpm \includegraphics[scale=0.35]{Link_S2_S1_in_S4.pdf} \epm 
={\tL}^{(S^2,S^1)}_{ \mu \sigma}
\end{matrix}$
& 1 & 
$
 \begin{matrix}
\ii \int \frac{ N_I}{2\pi}{B^I  \wedge \dd A^I}\\[2mm]
A^I \to A^I+ \dd g^I, \\[2mm]
N_I B^I  \to N_I B^I + \dd \eta^I.
 \end{matrix}
$ \\
\hline\\[-2mm]
$\begin{matrix}
Z \bpm \includegraphics[scale=0.35]{3+1D_Triple_link_mid_SpinHopfLink_Large_S2_123_l.pdf} \epm 
={\tL^{\text{Tri}}_{{\mu_3}, {\mu_2}, {\mu_1}}}\\
=Z[S^4; \text{Link[Spun[Hopf}[\mu_3,\mu_2]],\mu_1]] 
\end{matrix}$ &   
${\exp \big( \frac{2 \pi \ii p_{{IJK}}^{} }{ (N_{IJ} \cdot N_K  )   }    (a_I b_J )( c_K +d_K - [c_K+d_K  ]) \big)}$   & 
$\begin{matrix}
\ii \int 
\frac{ N_I}{2\pi}{B^I  \wedge \dd A^I} {{+}}  
\overset{}{\underset{{I,J}}{\sum}}
\frac{ N_I N_J \; p_{IJK}}{{(2 \pi)^2 } N_{IJ}}   
A^I \wedge A^J \wedge \dd A^K \\[2mm]
A^I  \to A^I + \dd g^I, \\[2mm]
 N_I B^I  \to N_I B^I  + \dd \eta^I +    \epsilon_{IJ}\frac{ N_I N_J \; p_{IJK}}{{2 \pi } N_{IJ}} d g^J \wedge A^K, \\[2mm] 
\text{here $K$ is fixed.} 
\end{matrix}$      
\\
\hline\\[-2mm]
$\begin{matrix}
Z \bpm \includegraphics[scale=0.35]{3+1D_SpinBorromean_mid_S2_1234.pdf} \epm\\
=Z[S^4; \text{Link[Spun[BR}[\mu_4 ,\mu_3, \mu_2]],\mu_1]];\\
\text{Also }  Z[T^4 \# S^2  \times S^2; \mu_4',\mu_3',\mu_2',\mu_1' ]
\end{matrix}$ & 
$\exp \big( \frac{2 \pi \ii p_{1234}}{ N_{1234} }  a_1 b_2 c_3 d_4 \big)$ & 
$\begin{matrix}
\ii \int  \frac{ N_I}{2\pi}{B^I  \wedge \dd A^I} + {{} c_{1234}} A^1 \wedge A^2 \wedge A^3 \wedge A^4 \\[2mm]
 A^I  \to A^I + \dd g^I, \\[2mm]
 N_I B^I  \to N_I B^I + d\eta^I -\pi {\tilde{c}}_{IJKL}  A^J  A^K g^L \\[2mm]
 + \pi {\tilde{c}}_{IJKL}  A^J  g^K d g^L 
- \frac{\pi}{3} {\tilde{c}}_{IJKL}  g^J  dg^K d g^L. 
\end{matrix}
$ \\
\hline
\end{tabular}
}
\caption{Examples of topological orders and their topological invariances in terms of our data in the spacetime dimension $d+1$D.
Here some explicit examples are given as Dijkgraaf-Witten twisted gauge theory \cite{Dijkgraaf:1989pz} 
with finite 
gauge group, such as $G=\Z_{N_1} \times \Z_{N_2} \times \Z_{N_3} \times \Z_{N_4} \times \dots$, 
although our quantum statistics data can be applied to more generic quantum systems without gauge or field theory description.
The first column shows the path integral form which encodes the braiding process of particles and strings in the spacetime.
In terms of spacetime picture, the path integral has nontrivial linkings of worldlines and worldsheets. The geometric Berry phases produced from
this adiabatic braiding process of particles and strings yield the measurable quantum statistics data. This data also serves as topological invariances for topological orders.
The second column shows the group-cohomology cocycle data $\omega$ as a certain partition-function solution of Dijkgraaf-Witten theory, where $\omega$ belongs to
the group-cohomology group, $\omega \in \cH^{d+1}[G,\R/\Z]=\cH^{d+1}[G,\mathrm{U}(1)]$.
The third column shows the proposed continuous low-energy field theory action form for these theories and their gauge transformations.
In 2+1D, $A$ and $B$ are 1-forms, while $g$ and $\eta$ are 0-forms.
In 3+1D, $B$ is a 2-form, $A$ and $\eta$ are 1-forms, while $g$ is a 0-form.
Here $I,J,K \in \{1,2,3,\dots\}$ belongs to the gauge subgroup indices, 
$N_{12\dots u} \equiv  \gcd(N_1,N_2, \dots, N_u)$ is defined as the greatest common divisor (gcd) of $N_1,N_2, \dots, N_u$.
Here $p_{IJ} \in \Z_{N_{IJ}}, p_{123} \in \Z_{N_{123}}, p_{IJK} \in \Z_{N_{IJK}}, p_{1234} \in \Z_{N_{1234}}$ are integer coefficients. The $c_{IJ}, c_{123}, c_{IJK}, c_{1234}$ are quantized coefficients labeling distinct topological gauge theories, where
$c_{12}=\frac{1}{(2 \pi)} \frac{N_1 N_2\;
p_{12 }}{N_{12}}$, $c_{123}=\frac{1}{(2 \pi)^2 } \frac{N_1 N_2 N_3\;
p_{123}}{N_{123}}$,  $c_{1234}=\frac{1}{(2 \pi)^3}  \frac{N_1 N_2
N_3 N_4\; p_{1234 }}{N_{1234}}$.
Be aware that we define
both $p_{IJ \dots}$ and $c_{IJ \dots}$ as constants with \emph{fixed-indices} ${I,J, \dots}$ without summing over those indices; 
while we additionally define
${\tilde{c}}_{IJ \dots} \equiv \epsilon_{IJ \dots} c_{12 \dots}$ with the $\epsilon_{IJ \dots}= \pm 1$ as an anti-symmetric Levi-Civita alternating tensor where
${I,J, \dots}$ are \emph{free indices} needed to be Einstein-summed over, but $c_{12 \dots}$ is fixed.
The lower and upper indices need to be summed-over, for example $\int \frac{ N_I}{2\pi}{B^I  \wedge \dd A^I}$ means that
$\int \overset{s}{\underset{{I=1}}{\sum}}  \frac{ N_I}{2\pi}{B^I  \wedge \dd A^I}$ where the value of $s$ depends on the total number $s$ of gauge subgroups $G=\prod_i^{s} \Z_{N_i}$. 
The quantization labelings are described and derived in \cite{{Wang:2014oya},{JuvenSPT1}}.
See the explicit verification of our proposed link invariants
from continuum TQFT formulations in \cite{Putrov2016qdo1612.09298}.
}
\label{table:TQFT}
}
\end{table}
\end{center}

\restoregeometry

The relevant field theories are also discussed in Ref. \cite{Kapustin:2014zva, JuvenSPT1, Gaiotto:2014kfa, Gu:2015lfa, Ye:2015eba, He2016xpiYunqin1608.05393, Wang2018iwzQR-Gu1810.13428}, 
here we systematically summarize and claim
the field theories in Table  \ref{table:TQFT}  third column indeed describe the low energy TQFTs of Dijkgraaf-Witten theory.  
It can be checked that the continuum TQFT formulation can indeed match to 
Dijkgraaf-Witten theory  \cite{Dijkgraaf:1989pz} and 
its discrete cochain formulation \cite{deWildPropitius1995hk9511201}, \cite{Wang:2014oya}, \cite{Wen2016cij1612.01418, Huang2017jxaZheyanWan1703.03266}.

Readers can find the similarity of our Table \ref{table:TQFT} and Ref.~\cite{Putrov2016qdo1612.09298}'s Table I.
However, we emphasize that the derivations and the logic of our present work and Ref.~\cite{Putrov2016qdo1612.09298} are somehow opposite.
\begin{enumerate}[label=\textcolor{blue}{\Roman*.}, ref={\Roman*},leftmargin=*]
\item 
Our present work \cite{JWangthesis, Wang2016qhf1602.05951} starts from the following inputs:
\begin{multline}
\text{(1) Spacetime topology, geometric topology and surgery properties}\\
\text{$\overset{\text{derive}}{\longrightarrow}$ (2) quantum surgery formulas
}\\
\text{$\overset{\text{derive}}{\longrightarrow}$ (3) possible link invariants and spacetime braiding process} \\
\text{$\overset{\text{detect}}{\longrightarrow}$ (4) group cohomology cocycles and TQFTs},
\end{multline}
which is going from the inputs of purely quantum mechanics and mathematical geometric topology, to obtain the left-hand-side (LHS) of Table \ref{table:TQFT}, then 
to obtain the middle, then the right-hand-side (RHS) of Table \ref{table:TQFT}. 

\item 
Ref.~\cite{Putrov2016qdo1612.09298} starts from the following inputs of:
\begin{multline}
\text{(4) group cohomology cocycles and continuum TQFTs}\\ 
 \text{$\overset{\text{derive}}{\longrightarrow}$ (3) link invariants and spacetime braiding process
}.
\end{multline}
which is logically going from the opposite direction. 
\end{enumerate}

Further recently, the formulations of continuum TQFTs or discrete cochain TQFTs have also been generalized, from
\begin{enumerate}[label=\textcolor{blue}{\arabic*.}, ref={\arabic*},leftmargin=*]
\item 
TQFTs for the bosonic Dijkgraaf-Witten theory (\cite{Putrov2016qdo1612.09298} and References therein), \\[2mm]
to
\item Fermionic version of TQFT gauge theory: Fermionic finite-group gauge theory and spin-TQFT, and their braiding statistics or topological link invariants \cite{Cheng2017ftw1705.08911,
Wang2018edf1801.05416, Guo2018vijGOPWW1812.11959} \\[2mm]
and
\item Higher-gauge theory as TQFTs, see some selective examples in \cite{Kapustin2013qsaThorngren1308.2926, Kapustin2013uxaThorngren1309.4721}, \cite{Gaiotto:2014kfa}, \cite{Putrov2016qdo1612.09298}, \cite{Chan2017eovYe1703.01926, Delcamp2018wlbTiwari1802.10104, Zhu2018kzd1808.09394, Wan2018bnsWW1812.11967, Delcamp2019fdpTiwari1901.02249}. 
\end{enumerate}

\section{Conclusion}

\subsection{Comparison to Previous Works}

It is known that the quantum statistics of particles in 2+1D begets 
anyons, beyond the familiar statistics of bosons and fermions, while Verlinde formula \cite{Verlinde:1988sn}
plays a pivotal role to dictate the consistent anyon statistics.
In this work, we derive a set of quantum surgery formulas analogous to Verlinde's constraining the fusion and braiding quantum statistics of 
 excitations of anyonic particle and anyonic string in 3+1D.
{We derive a set of fairly general quantum surgery formulas, which are also constraints of fusion and braiding data of topological orders.
We work out the explicit derivations of Eq.~(\ref{Z2Dcut}) in 2+1D, Eqs.~(\ref{eq:S2S1S1inS4}) and (\ref{eq:S1S2S2inS4}) in 3+1D, 
and then later we also derive Eq.~(\ref{eq:Spin[HopfLink]S4}) in 3+1D step by step.
We also derive the fusion constraint Eq.(\ref{eq:FS1FT2}) explicitly.}

A further advancement of our work, comparing to the pioneer work of Witten Ref.~\cite{Witten:1988hf} on 2+1D Chern-Simons gauge theory, 
is that we apply the surgery idea to generic 2+1D and 3+1D topological orders
without assuming quantum field theory (QFT) or gauge theory description. 
Although many lattice-regularized topological orders happen to have TQFT descriptions at low energy, 
we may not know which topological order derives which TQFT easily. 
Instead we simply use 
quantum amplitudes written in the bra and ket (over-)complete bases, obtained
from inserting worldline/sheet operators along the cycles of non-trivial homology group generators of a spacetime submanifold, 
to 
cut and glue to the desired path integrals. 
Consequently our approach, without the necessity of any QFT description, can be powerful to describe more generic quantum systems.\footnote{Namely, 
in our formulation and in our present derivation of quantum surgery formulas, 
we do \emph{not} require the second quantization  description of a quantum theory such as a QFT, which is more subtle to be formulated rigorously and mathematically.
Instead we only require the first quantization description of a quantum theory via the standard mathematically well-defined standard \emph{quantum mechanics}.
}
While our result is originally based on 
studying specific examples of TQFT (such as Dijkgraaf-Witten gauge theory \cite{Dijkgraaf:1989pz}), 
we formulate the 
data without using QFT. 
We have incorporated the necessary generic quantum statistic data and new constraints 
to characterize some 3+1D topological orders (including Dijkgraaf-Witten's),
we will leave the issue of their sufficiency and completeness 
for future work. 
Formally, our approach can be applied to any spacetime dimensions.


\subsection{Physics and Laboratory Realization, and Future Directions}

Now we comment about the more physical and laboratory realization of implementing the cut-and-glue surgery procedure we discussed earlier.
The matrix obtained from quantum wavefunction-overlap between different 
ground states of Hilbert space of a quantum matter,\footnote{
Preferably, the following discussion is rigorous and well-defined, 
when we limit our discussion such that quantum matter is topologically ordered, the so-called 
topological order.
} 
forms a representation of the mapping class group (MCG) of the real space manifold ($M_{\text{space}}$) where the quantum matter resides (See \cite{Moradi2014kla1401.0518}, \cite{Moradi:2014cfa, Wang:2014oya}, and references therein). This matrix can be written as:
\bea
&&\varphi_{({\sigma', \mu',\dots}),({\sigma, \mu,\dots})} \equiv
\langle \psi_{\sigma', \mu',\dots} | \hat \varphi  | \psi_{\sigma, \mu,\dots} \rangle,
\\
&&   \hat \varphi \in \text{MCG}(M_{\text{space}}), \quad \quad | \psi_{\sigma, \mu,\dots} \rangle \in \cH,
\\
&& \text{rank}{(\varphi_{({\sigma', \mu',\dots}),({\sigma, \mu,\dots})})} = \text{GSD}_{M_{\text{space}}}=\dim(\cH) = |{Z}({M_{\text{space}} \times S^1})|.
\eea
 It is worthwhile to mention that for 2+1D topological order, 
 one can construct group elements of the mapping class group of a genus g Riemann 2-surface $\cM_g$, 
 obtained via a series of projections on selecting the quasi-excitation sectors (i.e., projection by selecting the ground state sector $ | \psi_{\sigma, \mu,\dots} \rangle$), 
 along at least above a certain number of mutually intersecting non-contractible cycles on the  $\cM_g$ \cite{Barkeshli2016ivz1602.01093}.
 It is noted that\\ 
 (i) A genus g Riemann 2-surface $\cM_g$ can be realized 
 in a 2D planar geometries via multi-layers or folded constructions of the desired topological orders, 
with the appropriate types of gapped boundaries designed.\\
 (ii) The projections on selecting the quasi-excitation sectors (named topological charge projections in \cite{Barkeshli2016ivz1602.01093}) 
 can be implemented as the adiabatic unitary deformations 
 by tuning microscopic parameters of the quantum system locally 
 (at the lattice scale or at the ultraviolet high energy locally). \\
The result suggests that, in 2+1D,
we could potentially implement and realize 
the quantum representation of modular transformations or MCG($\cM_g$) of topological orders in physical systems.\\

Future Directions:
\begin{enumerate}[label=\textcolor{blue}{\arabic*.}, ref={\arabic*},leftmargin=*]
\item 
It will be interesting to study the 3+1D story analogous to what \cite{Barkeshli2016ivz1602.01093} has done for the ``mapping class group representation realization'' in 2+1D,
given that our present and earlier work indicate exotic braiding, fusion and ``quantum surgeries'' happen in 3+1D
\cite{JWangthesis, Wang2016qhf1602.05951}.

\item \label{direct2}
There is a recent interest concerning a constraint of fusion and half-braiding process\footnote{The half-braiding means that the braiding of the bulk excitations in 3d (2+1D)
is moving through a half-circle 1-worldline to the constrained 2d (1+1D) boundary.
} 
of boundary excitations of 2+1D topological orders (or 3d TQFTs),
coined boundary's defect Verlinde formula \cite{Shen2019wopHung1901.08285}. 
The approach relies on a tunneling matrix data $\mathcal{W}_{ia}$,
defined in \cite{Lan2014uaaLWW1408.6514}, which shows how an anyon $i$ in 3d TQFT Phase A
can decompose into a direct sum of a set of anyons $\oplus_{a} \mathcal{W}_{ia}a$ after tunneling to another 3d TQFT Phase B.
Ref.~\cite{Shen2019wopHung1901.08285} suggests a defect Verlinde formula (on 2d boundary of 3d TQFT)
based on the relation of data of:\\ 
$\bullet$ The fusion rules of the boundary excitations in 2d, and \\
$\bullet$ The half-braiding or half-linking modular data (a modified modular $\cS$ matrix) for 3d TQFT with 2d boundary.\\
It will be an illuminating future direction to derive an analogous story for 3d (2+1D) boundary of a 4d (3+1D) TQFT with an analogous boundary excitation quantum surgery formulas like ours (Eqs.~(\ref{eq:S2S1S1inS4}), (\ref{eq:S1S2S2inS4}) 
and (\ref{eq:Spin[HopfLink]S4}) in 4d).
Such a boundary's defect Verlinde formula may be helpful to constrain other physical observables of systems of gapped boundary with defects, such as the 
\emph{boundary topological degeneracy} \cite{Wang2012am1212.4863,Kapustin2013nva1306.4254,Hung2014tbaWan1408.0014}
\cite{Lan2014uaaLWW1408.6514}.  
Moreover, 
by ``the bulk-boundary 3d-2d correspondence,'' we see that\\
$\clubsuit$ The 2d boundaries/interfaces of 3d TQFT systems can be regarded as the 2d defects surfaces in 3d TQFT;\\
via ``dimensional reductions (lowering one dimension)'' thus the above discussion intimately relates to\\
$\clubsuit$ The 1d boundaries/interfaces of 2d CFT, or the 1d defect lines in 2d CFT. 
The later direction of 2d-1d system had generated various interests in the past 
\cite{Kramers1941knPhysRev, Cardy1986gwNPB, Oshikawa1996dj9612187, Petkova2000ipZuber0011021, Frohlich2004ef0404051, Frohlich2006ch0607247} and fairly recently \cite{Davydov2010rmLiangKong1004.4725, Gaiotto2014lma1404.0332, Chang2018iay1802.04445}.\\
Topological defects in higher dimensions viewed as the gapped interfaces/boundaries in any dimensions of group cohomology theory is explored recently in \cite{Wang2017locWWW1705.06728}. So Ref.~\cite{Wang2017locWWW1705.06728}'s study may play a helpful guiding role along this direction.
Possible outcomes along this direction may lead us to understand:\\
 $\spadesuit$ Relation between \emph{Verlinde formula} of 3d TQFT/2d CFT and the \emph{defect Verlinde formula} \cite{Shen2019wopHung1901.08285} (on 2d boundary of 3d TQFT, or 1d defect line of 2d CFT).\\
 to\\
 $\spadesuit$ Relation between our \emph{quantum surgery formulas} \cite{JWangthesis, Wang2016qhf1602.05951}
 of 4d TQFT/3d CFT and the \emph{defect analogous of these quantum surgery formulas} (on 3d boundary of 4d TQFT; or 1d defect 
 line or 2d defect surface of 3d CFT).

 \item
It will be interesting to study the analogous Verlinde formula constraints for 2+1D boundary states of 3+1D bulk systems 
(thus relates to the direction \ref{direct2}), such as highly-entangled gapless modes,
3d conformal field theories (CFT) and 3d anomalies.\footnote{The 't Hooft anomalies of global symmetries can be realized as the non-onsite-ness of
the global symmetry acting on the boundary states. The gauged version of the ``whole'' bulk-and-anomalous-boundary system also has interesting physical consequences.
The interpretation of anomalies along the group cohomology theory can be found in \cite{Wang2014tia1403.5256, Kapustin:2014zva, Wang2017locWWW1705.06728} and References therein.
} 
For example, one can start from exploring the bulk-boundary correspondence between TQFT and CFT;
such as
for 3d TQFT and 2d CFT (\cite{Witten:1988hf} and the group-cohomology version of correspondence \cite{Wang2014tia1403.5256}), 
and for 4d TQFT and 3d CFT \cite{2015arXiv150904266C, 2015arXiv151209111W}, 
of some bulk quantum systems.
The set of consistent quantum surgery formulas 
we derive may lead to an alternative effective way to \emph{bootstrap} \cite{Polyakov:1974gs,Ferrara:1973yt} 3+1D topological states of matter and 2+1D CFT.\\

\end{enumerate}

\noindent
{\bf Note added}:
The formalism and some results discussed in this work have been partially reported in the first author's Ph.D. thesis \cite{JWangthesis} and in \cite{Wang2016qhf1602.05951}. 
Readers may refer to Ref.~\cite{JWangthesis} for other discussions.
Other related aspects of research will also be reported in upcoming work \cite{WWY-Gompf}, \cite{WY} and  \cite{WZ}.


\section{Acknowledgements} 

We are indebted to Clifford Taubes for many generous helps at various stages during the development of this work. 
JW is grateful to 
Ronald Fintushel, 
Robert Gompf, Allen Hatcher, Shenghan Jiang, 
Ronald Stern, Andras Stipsicz, 
and
Edward Witten 
for helpful comments, 
and to colleagues at 
Harvard University for discussions.
JW is supported by
NSF Grant PHY-1606531.  
XGW is partially supported by NSF grant DMR-1506475 and DMS-1664412. 
This work is also supported by 
NSF Grant DMS-1607871 ``Analysis, Geometry and Mathematical Physics'' 
and Center for Mathematical Sciences and Applications at Harvard University.

\bibliographystyle{Yang-Mills.bst}
\bibliography{Top-faMCG_ref_new,cross,all}

\end{document}